\documentclass[%
 aip,
 amsmath,amssymb,
 reprint,%
]{revtex4-1}

\usepackage{graphicx}% Include figure files
\usepackage{dcolumn}% Align table columns on decimal point
\usepackage{bm}% bold math
\usepackage[utf8]{inputenc}
\usepackage[T1]{fontenc}
\usepackage{mathptmx}
\usepackage{etoolbox}

\makeatletter
\def\@email#1#2{%
 \endgroup
 \patchcmd{\titleblock@produce}
  {\frontmatter@RRAPformat}
  {\frontmatter@RRAPformat{\produce@RRAP{*#1\href{mailto:#2}{#2}}}\frontmatter@RRAPformat}
  {}{}
}%
\makeatother
\begin{document}

\preprint{AIP/123-QED}

\title{Topographic De-adhesion in the Viscoelastic Limit}
% Force line breaks with \\
% Force line breaks with \\
\author{Nhung Nguyen}
\affiliation{Department of Surgery, The University of Chicago, Chicago, IL, USA.}%Lines break automatically or can be forced with \\
\author{Eugenio Hamm Hahn}%
\affiliation{Facultad de Ciencia, Departamento de Física, Universidad de Santiago de Chile (USACH), Santiago, Chile.} 
\author{Sachin Velankar}
\affiliation{Department of Chemical Engineering, University of Pittsburgh, Pittsburgh, PA, USA.}
\author{Enrique Cerda}
\affiliation{Facultad de Ciencia, Departamento de Física, Universidad de Santiago de Chile (USACH), Santiago, Chile.}
\author{Luka Pocivavsek}
\affiliation{Department of Surgery, The University of Chicago, Chicago, IL, USA.}
\email{lpocivavsek@bsd.uchicago.edu.}
%\date{\today}% It is always \today, today,
             %  but any date may be explicitly specified

\begin{abstract}
The superiority of many natural surfaces at resisting soft, sticky biofoulants has inspired the integration of dynamic topography with mechanical instability to promote self-cleaning artificial surfaces. The physics behind this novel mechanism is currently limited to elastic biofoulants where surface energy, bending stiffness, and topographical wavelength are key factors. However, the viscoelastic nature of many biofoulants causes a complex interplay between these factors with time-dependent characteristics such as material softening and loading rate. Here, we enrich the current elastic theory of topographic de-adhesion using analytical and finite element models to elucidate the non-linear, time-dependent interaction of three physical, dimensionless parameters: biofoulant's stiffness reduction, product of relaxation time and loading rate, and the critical strain for short-term elastic de-adhesion. Theoretical predictions, in good agreement with numerical simulations, provide insight into tuning these control parameters to optimize surface renewal via topographic de-adhesion in the viscoelastic regime. 
\end{abstract}

\maketitle

\section*{Statement of significance}
Topography driven surface renewal is actively being applied to at-risk medical devices such as vascular grafts and catheters. A significant knowledge gap exists in the relevant parameter space controlling fracture at the biofoulant-material interface. In particular, the interlay of multiple non-linearities arising from surface geometry (wrinkling) and material response (viscoelasticity) combined with fracture pose a challenging problem. This paper explores the physics of dynamic surface topography on interfacial stability when multiple sources of energy dissipation co-exists: interfacial fracture and viscoelasticity. Our calculations, validated by numerical simulations, provide the target critical strain for topographic renewal of a viscoelastic foulant. Furthermore, we provide surface design specifications to optimize surface renewal for biofoulants that can immediately be used by biomedical engineers.

%\begin{quotation}
%The ``lead paragraph'' is encapsulated with the \LaTeX\ 
%\verb+quotation+ environment and is formatted as a single paragraph before the first section heading. 
%(The \verb+quotation+ environment reverts to its usual meaning after the first sectioning command.) 
%Note that numbered references are allowed in the lead paragraph.
%%
%The lead paragraph will only be found in an article being prepared for the journal \textit{Chaos}.
%\end{quotation}

\section{\label{sec:level1}Introduction}
\begin{figure*}
\centering
\includegraphics[width=1\textwidth]{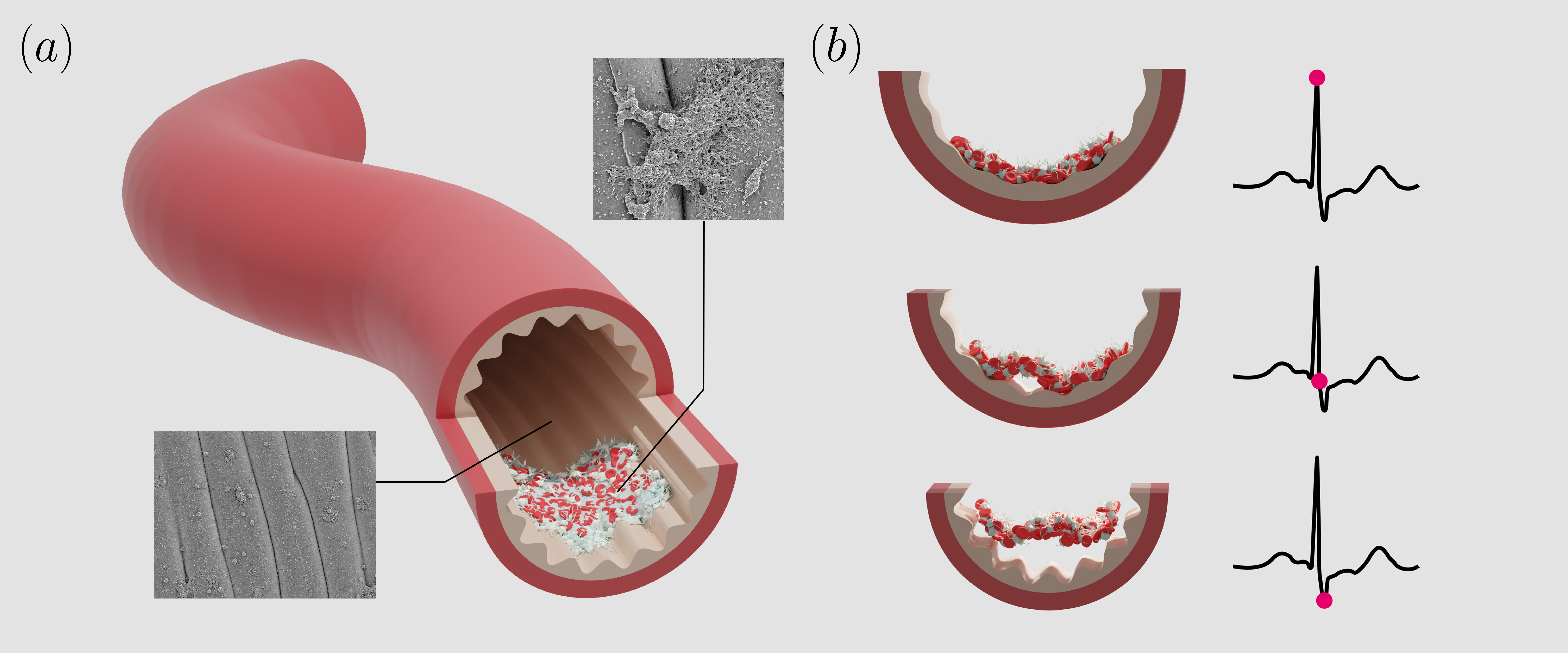}
\caption{Topography-driven delamination motivated by dynamically actuated wrinkled surface in artery. (a) A three dimensional view of biofoulants attached to the arterial luminal wrinkles. (b) Cross-sectional views of the detachment process of biofoulants from the wrinkled topography at different levels of actuated pulse pressure through the cardiac cycle illustrated in blue on the right. During systole (top), the high pressure distends the artery and flattens the luminal wrinkling. As the cardiac cycle progresses, the arterial pressure drops gradually until the diastolic pressure (bottom) is reached. As the pressure decreases, the artery contracts and the amplitude of the wrinkles grow. The cardiac cycle then repeats when the left ventricle again contracts pumping blood into the arterial system. Current attempts with experiment and computational modeling focus on elastic behaviors of the biofoulant, and show a scaling dependence of the critical surface curvature on the surface energy, bending stiffness, and topographic geometrical wavelength \cite{Luka2018}.}
\label{ArterialWrinkle}
\end{figure*}

Many natural surfaces including airways, arteries, and intestines have dynamically actuated geometries \cite{Li2014, Bixler2012, Genzer2006, Pocivavsek2009, Russel2002, Shivapooja2013, Levering2014, Luka2018, Luka2019,Tindall1988,Nandan2020}. Specifically, pulse pressure in arteries is believed to drive the luminal arterial geometry between low and high curvature states, generating an actuating topography \cite{ Luka2018, Luka2019,Tindall1988,Nandan2020}. Figure \ref{ArterialWrinkle} schematically demonstrates the change in arterial topography during the cardiac cycle,  with a flat luminal surface at systole (Figure \ref{ArterialWrinkle}b top) and a wrinkled surface at diastole 
%%%%%%%%%%%%%%% End of first page %%%%%%%%%%%%%%%%%%%%%
\maketitle
\noindent  (Figure \ref{ArterialWrinkle}b bottom). The existence of these states has been experimentally validated ex-vivo in fresh arterial segments in a previous work \cite{Nguyen2020}. Such topographic variations may play an important role in keeping these surfaces clean from biofouling on macroscopic scales and potentiate nature's multi-scale anti-fouling strategies at complex interfaces. Inspired by these natural phenomena, a new de-adhesion mechanism, using dynamic surface topography, was discovered: topography-driven delamination (Figure \ref{ArterialWrinkle}), where foulant deformation, driven by evolving surface curvature, generates an energy release mechanism by balancing elastic energy with adhesion strength \cite{Shivapooja2013, Levering2014, Luka2018, Luka2019, Nandan2020}. Thus far, topography-driven delamination has only been explored in the idealized elastic foulants, where all stored elastic energy is available to drive surface renewal once a critical surface curvature, $\kappa_c$, is reached \cite{Luka2018}. 

Biofoulants, such as platelets, thrombus, biofilms, and bacteria, are complex, dynamic materials. In order to integrate topographic anti-fouling strategies into biomaterials and medical devices with specific engineering design limits with given loads and achievable actuation strains, topography-driven delamination must be understood in the viscoelastic regime that more accurately represents the material properties of biological foulants \cite{Bixler2012, Luka2018, Hasan2015, Chen2011, Mao2009, Koh2010, Shaw2004}. Viscoelasticity allows a continuous softening of material properties, a dependence on loading history, and an intrinsic mechanism for dissipating elastic energy within a stressed material \cite{Christensen2003, Wineman2000}. Since topography-driven delamination is an energy release mechanism where available foulant elastic energy drives interfacial fracture, the existence of a competing dissipation mechanism, intrinsic to the material, enriches the problem substantially, compared to the elastic limit. 

\begin{figure*}
\centering
\includegraphics[width=1\textwidth]{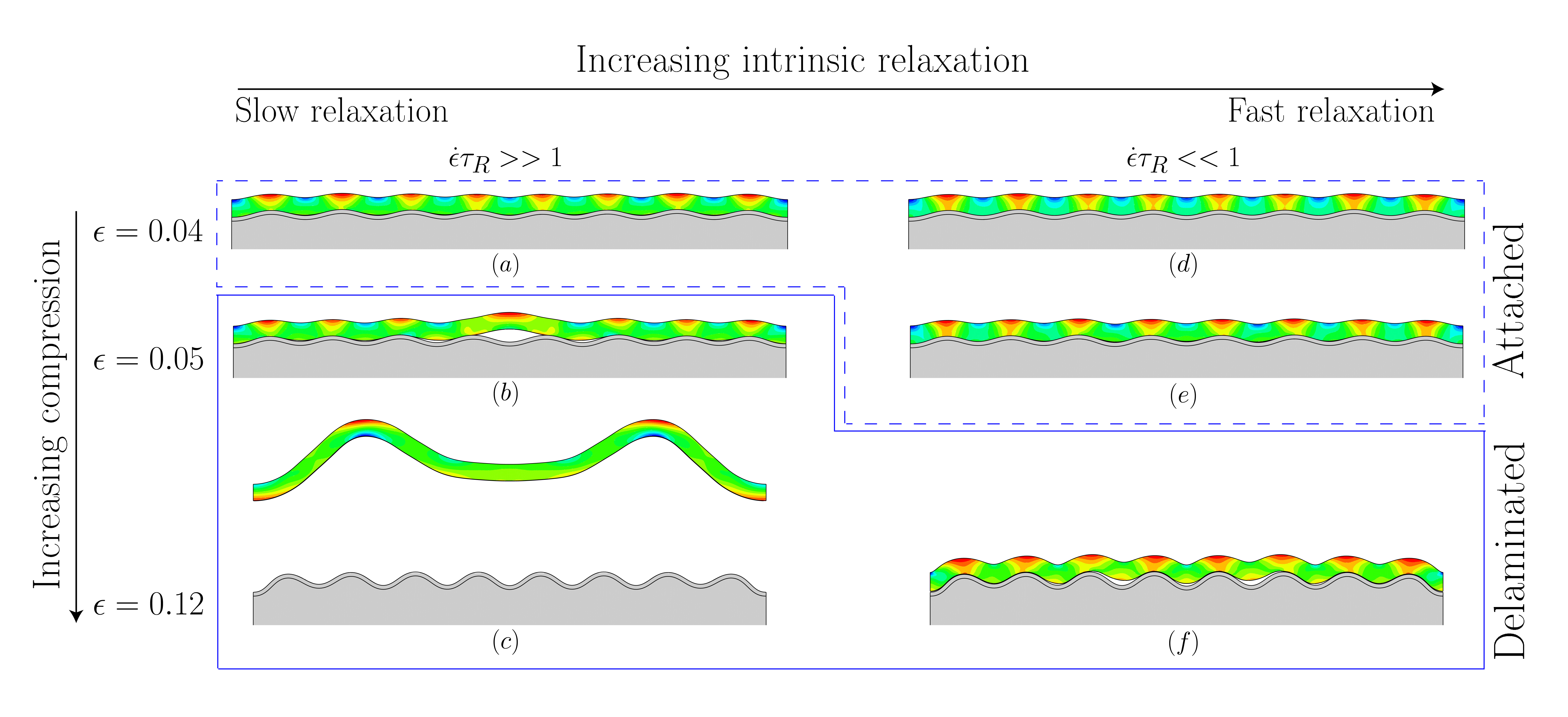}
\caption{Delamination processes under increasing applied strain $\epsilon$ of a slowly relaxing viscoelastic foulant layer (a, b, c) as compared to a fast relaxing viscoelastic foulant layer (d, e, f). The latter conforms more stably to the wrinkled topography and requires a higher strain $\epsilon = 0.12$ (Figure f) than the former $\epsilon = 0.05$ (Figure b) to cause its detachment.}
\label{DetachProcess}
\end{figure*}

Figure \ref{DetachProcess} shows a striking difference between the detachment process of an elastic (or equivalently, a viscoelastic foulant layer with large intrinsic relaxation time $\tau_R$ relative to the intrinsic loading rate $\dot{\epsilon}$) and a viscoelastic foulant layer with a short relaxation time scale (see also Supplementary video S1). In both cases, the foulant layers initially follow the wrinkled surface, which is the source of changing curvature $\kappa=A/\lambda^2$, where $A$ is wrinkle amplitude and $\lambda$ is wrinkle wavelength. They ultimately delaminate, however the one with fast intrinsic relaxation, $\dot{\epsilon}\tau_R \ll 1$, remains attached until higher critical curvatures (critical strains). These examples show that the presence of significant viscoelasticity, on the time scale of loading, stabilizes the foulant layer/substrate interface. Of note, the nominal compressive critical strain more than doubles from $\sim 5\%$ to $\sim12\%$ in these two cases. 

For optimal design of self-cleaning surfaces in contact with viscoelastic biofoulants, the interaction of the time-dependent characteristics of the viscoelastic foulant layer with the geometry set by the wrinkled topography and the surface energy needs to be understood, especially for dynamic topography-driven anti-fouling strategies in biomaterials and medical devices \cite{Shivapooja2013, Levering2014, Luka2019, Nandan2020}. In this paper, we tackle the complex interaction between foulant layer's viscoelasticity, rate of topographic changes, mechanical instability, and surface adhesion in the wrinkle-induced delamination model \cite{Luka2018}. Utilizing analytical methods based on energy minimization \cite{Williams1994,Hutchinson1992,Hutchinson2017,Vella2009, Luka2018, Oz2018, Paul2012, Chai1981, Ravichandran1994, Christensen2003, Roger2013} and finite element analysis with cohesive zone modeling \cite{Mei2011, Thouless2007,Turon2007, Xie2006, Heinrich2012, Paul2012, NN2016, Lin2019, NN2017, Thouless2019, Luka2018}, we reduce the parameter space that controls the delamination of a thin, viscoelastic foulant layer from a wrinkled surface to three physical dimensionless parameters: magnitude of foulant layer's relaxation, rate of material relaxation relative to loading rate, and the critical strain for delamination in the short-term (instantaneous elastic) limit. The last parameter incorporates the surface energy, the bending stiffness of the foulant layer, and the surface geometry via the wrinkle wavelength. Our analytical model for viscoelastic topographic de-adhesion is able to collapse numerical data from a large range of the dimensionless parameter space. Our analysis provides insight into viscoelastic delamination for geometrically non-linear interfaces that inevitably exist in biological systems and for artificial materials incorporating topography as self-cleaning strategies \cite{Shivapooja2013, Levering2014, Luka2018, Luka2019, Nandan2020}. 
\section{Results}
\subsection{Scaling Analysis}
\begin{figure*}
\centering
\includegraphics[width=1\textwidth]{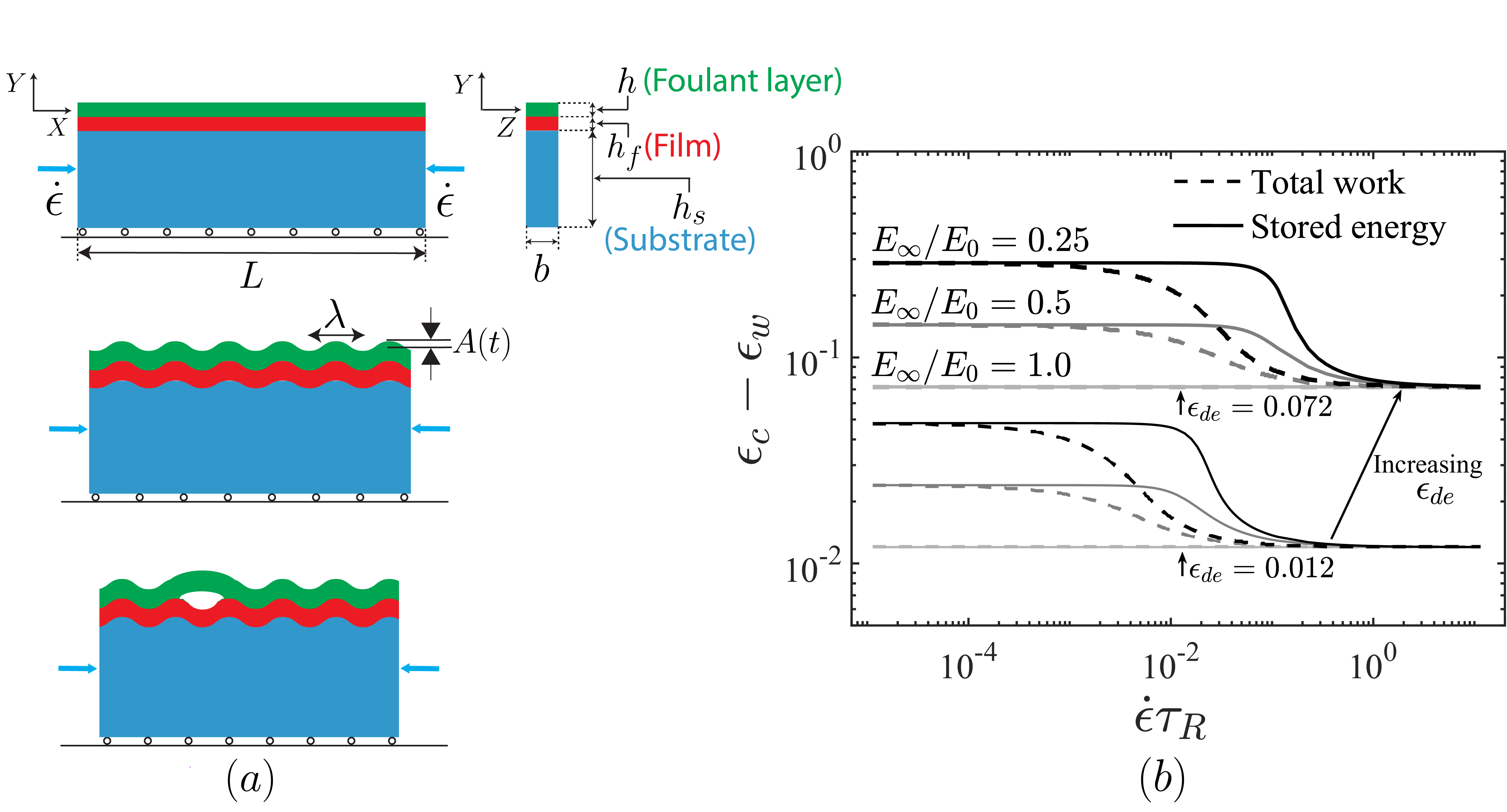}
\caption{Compression of a viscoelastic foulant layer attached to a bilayer for studying wrinkle-induced delamination. (a) Schematics of a thin viscoelastic foulant layer attached to a bilayer system composed of a thin film on top of a substrate subjected to increasing compression. The bilayer wrinkles which induces a wrinkle pattern on the foulant layer. As the wrinkle curvature increases $\kappa = A/\lambda^2$, the foulant layer starts to de-adhere from the bilayer. Here, $A$ and $\lambda$ are the wrinkle amplitude and wavelength. (b) Energy balance approach shows the dependence of the critical strain on three controlling dimensionless parameters. Two representative values of $\epsilon_{de}$ and three representative values of $E_\infty/E_0$ are selected for plotting here. Note that when $E_\infty/E_0=1.0$, the result of the elastic foulant layer \cite{Luka2018} is reproduced as shown by the horizontal, flat curves. In these cases, the critical strain is equal to $\epsilon_{de}$ and is independent of $\dot{\epsilon}\tau_R$.}
\label{SchematicEApproach}
\end{figure*}

A viscoelastic foulant layer (adherent biofoulant) of thickness $h$ is attached to the surface of an elastic bilayer designed to mimic the multi-layered structure of the arterial wall \cite{Luka2018,Nguyen2020}. The resultant tri-layer system is subjected to compression as shown in Figure \ref{SchematicEApproach}a. Under increasing applied compression on the two ends and assuming plane strain in the orthogonal z direction, the bilayer composed of a stiff film of thickness $h_f$ attached to a soft substrate of thickness $h_s \gg h_f$, undergoes wrinkling with increasing wrinkle amplitudes as described previously \cite{Genzer2006, Luka2018, Allen1969, Bowden1998, Pocivavsek2008, Sun2012, Cao2012, Cerda2003}. Specifically, the critical strain for wrinkle onset is set by the mismatch stiffness between the film and the substrate: $\epsilon_w \sim \left(E_s/E_f\right)^{2/3}$ where $E_s$ and $E_f$ are the substrate and the film moduli, respectively. Surface topography is set by its curvature $\kappa=A/\lambda^2$, which depends on wrinkle wavelength $\lambda \sim h_f \left(E_f/E_s\right)^{1/3}$ and amplitude $A \sim \lambda \sqrt{\epsilon-\epsilon_w}$, where $\epsilon$ is the applied nominal strain. This wrinkle pattern is not affected by the foulant layer because the foulant layer is significantly softer than the constituents of the bilayer \cite{Luka2018}. To reduce the effect of the pre-wrinkled state on the subsequent delamination process, we study the regime where the wrinkle onset strain $\epsilon_w$ is small, thereby $A \sim \lambda \sqrt{\epsilon}$. As a critical curvature $\kappa_c \sim A_c/\lambda^2 \sim \sqrt{\epsilon_c}/\lambda$ is reached, the viscoelastic foulant layer starts to de-adhere from the bilayer surface. In the elastic limit for the foulant layer solved by Pocivavsek et al. \cite{Luka2018}, no time-dependent quantity enters the solution $\epsilon_c = \epsilon_{de} = a \lambda^2 G/B$, where $G$ is the adhesion energy, $B=Eh^3/12(1-\nu^2)$ is the bending stiffness of the elastic foulant layer ($\nu$ is the Poisson's ratio of the foulant layer), and $a$ is a numerical prefactor. In the current case of viscoelastic foulant layer, the intrinsic relaxation time of the material interacts with the time scale set by the rate of topographic loading to give rise to a time-dependent delamination condition \cite{Bazant}. Here, for the simplest viscoelastic foulant layer, the relaxation is described using a single term Prony series: $E(t)=E_\infty+(E_0-E_\infty)e^{-t/\tau_R}$, where $\tau_R$ is the relaxation time and $E_\infty \le E_0$ are the long-term and instantaneous elastic stiffnesses of the foulant layer, respectively. A mechanical analog for this representation is a Maxwell model with two springs of stiffnesses $E_\infty$ and $E_0-E_\infty$ and a dashpot to dissipate energy (see Supplementary Appendix 1). With this decaying function of the modulus, a simple mathematical extension of the result of Pocivavsek et al. \cite{Luka2018} for compression applied at a constant rate $\epsilon=\dot{\epsilon}t$ is to substitute $E(t)$ into the scaling law derived in the elastic limit, which leads to the identity: 
\begin{equation}
\epsilon_c  = \frac{a\lambda^2 G/B_0}{1-\beta \left(1-e^{-\epsilon_c/(\dot{\epsilon}\tau_R)}\right)}.
\end{equation}
In terms of time scales, this equation is written as: 
\begin{eqnarray}
t_c/\tau_R = \frac{\tau_{de}/\tau_R}{\left[1-\beta(1-e^{-t_c/\tau_R})\right]},
\label{TimeScales}
\end{eqnarray}
where $\epsilon_c=\dot{\epsilon}t_c$ and $t_c$ are the critical strain and delamination time, respectively; $\beta=(B_0-B_\infty)/B_0=(E_0-E_\infty)/E_0$ is the fraction of the stiffness reduction in the foulant layer, with $B_0=E_0 h^3/12(1-\nu^2)$ and $B_\infty=E_\infty h^3/12(1-\nu^2)$ providing bending stiffnesses of the foulant layer at the instantaneous and long-term elastic limits, respectively; and $\tau_{de}= \epsilon_{de}/\dot{\epsilon}$ is the characteristic time for delamination in the instantaneous elastic limit. 

Three physical dimensionless parameters emerge from these equations: $E_\infty/E_0$ (magnitude of the foulant layer's relaxation), $\,\dot{\epsilon}\tau_R$ (rate of material relaxation relative to loading rate, which is similar to the Weissenberg number used by rheologists to quantify viscoelastic effects \cite{Poole2012}), and $\epsilon_{de}$ (critical delamination strain at the instantaneous elastic limit which depends on wrinkle wavelength, adhesion strength, and the instantaneous elastic modulus of the foulant layer as quantitatively described by the elastic model of Pocivavsek et al. \cite{Luka2018}). The ratio between the last two parameters $\tau_{de}/\tau_R$ is a control parameter for the system: when $\tau_{de}/\tau_R \ll 1$ we expect from the balance in Equation \ref{TimeScales} that  $t_c/\tau_R\ll 1$. It implies that the denominator is approximately equal to one such that $t_c\approx \tau_{de}$. Similarly, when $\tau_{de}/\tau_R \gg 1$ the same balance gives $t_c/\tau_R\gg 1$ and the result $t_c\approx \tau_{de}/(1-\beta)=a \lambda^2G/(\dot{\epsilon}B_\infty)$. Thus, delamination is controlled by the bending stiffness $B_0$ for $\tau_{de}/\tau_R\ll1$ and the opposite limit corresponds to delamination controlled by $B_\infty$. 

\subsection{Energy Analysis}
Though the above mathematical approach indicates the presence of important dimensionless parameters controlling this complex viscoelastic system and their non-linear coupling, to obtain a physical mechanism driving delamination in this system, we employ an analysis based on energy minimization. In the case of a purely elastic foulant layer without delamination, the work performed by the wrinkled surface generates stored elastic energy in the foulant layer $\dot{W}=\dot{U}$, where the total work injected into the system is $\dot{W}=\int_V \sigma\dot{\epsilon}dV$ and $\sigma=E\epsilon$. In the case of a viscoelastic foulant layer, part of the total work is stored and the other part is dissipated through viscous relaxation: $\dot{W}=\dot{U}+\dot{D}$. Here again $\dot{W}=\int_V\sigma\dot{\epsilon}dV$ but the stress and strain are divided in two parts: $\sigma=E_\infty(\epsilon^e+\epsilon^v)+(E_0-E_\infty)\epsilon^e$ and $\epsilon=\epsilon^e+\epsilon^v$, with $\epsilon^e$ being the elastic strain and $\epsilon^v$ being the viscous strain. Using a mechanical analog of linear viscoelasticity (see Supplementary Appendix 1), the stored elastic energy of the foulant layer is derived as: 
\begin{eqnarray}
U&=&\frac{E_\infty}{2}\int_V dV (\epsilon^e+\epsilon^v)^2 \nonumber\\
&+& \frac{E_0-E_\infty}{2}\int_V dV {(\epsilon^e)}^2.
\end{eqnarray}
Note that for $t\ll\tau_R$ no viscous dissipation has yet occurred as no viscous strain has had time to develop, thus $D=0$, $\epsilon^v=0$, and $U=(E_0/2) \int_V dV {(\epsilon^e)}^2$.  On the other hand, for $t\gg\tau_R$ the material has exhausted all possible sources of viscous dissipation, thus the foulant layer's elastic strain $\epsilon^e=0$ and $U=(E_\infty/2) \int_V dV {(\epsilon^v)}^2$. Since the total strain $\epsilon=\epsilon^e+\epsilon^v$, we see in the two limits the available energy is simply that of the purely elastic case with the two moduli $E=E_0$ and $E=E_\infty$ in the single term Prony series.  In other words, a physical interpretation of the stored energy is the elastic energy stored in the two elastic springs of the Maxwell model \cite{Christensen2003}. At the instantaneous limit $t\ll\tau_R$, $U$ and $W$ reduce to the bending energy for the elastic case with modulus $E_0$. At the long-term limit $t\gg\tau_R$, $U$ and $W$ have the same form of the elastic bending energy but with modulus $E_\infty$. However, in the intermediate regime between these two limits, the difference between $U$ and $W$ arises due to active viscous dissipation in this system (see detailed derivations in Supplementary Appendix 2). It is in this active regime, where two time-dependent mechanisms interact, that the analysis becomes interesting and complex. 

To drive fracture, a balance between the energy available in the system and the surface energy is required. For a conservative system, $dW=dU$ and a balance between the stored energy and the surface energy $d(U+U_S)=d(W+U_S)=0$, where $U_S=Gl$ and $l$ is the crack length, is widely used to study fracture \cite{Vella2009, Luka2018, Oz2018,Paul2012, Chai1981}. However, for a dissipative system, due to the presence of the material's dissipation energy, two arguments are presented in the literature for the source of energy to predict fracture. One approach balances the total work $W$ against the surface energy as a condition for crack propagation \cite{Roger2013,Christensen2003}: $d(W+U_S)=0$. A second approach assumes that only the stored energy $U$ is the amount of energy available for fracture \cite{Ravichandran1994}: $d(U+U_S)=0$. Note that both approaches are equivalent for a conservative system. The correct energy is debatable and more accurate experimental data and understanding of other possible sources of energies that might play a role in real systems such as dynamic effects  are needed to resolve this conflict. Yet, these two approaches provide insights into two limiting cases for the energy available for driving fracture in a non-conservative, viscoelastic system. In particular, the first one considers the largest possible amount of energy while the second one considers the smallest possible amount of energy available to overcome surface energy and cause fracture to occur. Thus, in order to physically understand the emergence and coupling of the three dimensionless parameters and how they govern the delamination of the viscoelastic foulant layer from the wrinkled surface, we use both approaches in our analytical models and compare them with the results obtained from numerical simulations. 

Specifically, from the imposed wrinkled topography, the height of the mid-plane of the foulant layer is $h_m(x,t)=A(t)\sin(kx)$, where $t=\epsilon/\dot{\epsilon}$ and $k=2 \pi/\lambda$ is the wave number. The curvature is computed as $\kappa=\partial^2_{x} h_m(x,t)=-k^2A(t)\sin(kx)$ giving the bending strain $\gamma=\kappa y$ at a point located at $0\le x\le L$ along the  length $L$ and at a distance of $y$ to the neutral axis (the mid-plane) of the thin foulant layer. Thus, the strain in the foulant layer is computed as $\gamma= -yk^2A(t)\sin(kx)$ (see Supplementary Appendix 2). Neglecting the small compressive strain prior to buckling, the viscous strain in the foulant layer becomes $\epsilon_{fl}^{v} = \left(1/\tau_R\right)\int_0^t \gamma(\tau)e^{-(t-\tau)/\tau_R}d\tau$ and the elastic strain is $\gamma^{e}=\gamma-\gamma^{v}$. Both approaches lead to non-linear, time-dependent equations whose solutions give the critical time (equivalently critical strain) to trigger delamination: 
\begin{eqnarray}
t_c/\tau_R = \frac{\tau_{de}/\tau_R}{\left[1-\beta P_{\pm}(t_c/\tau_R)\right]},
\label{Eqn4}
\end{eqnarray}
where $P_{-}(z)=\frac{1}{z}\int_0^z dw \frac{F(w)}{\sqrt{w}}$ and $P_{+}(z)=1-\left(1-\frac{F(z)}{\sqrt{z}}\right)^2$ are obtained using the total work $W$ and stored energy $U$, respectively, with $F(z)=\int_0^w \sqrt{y}e^{-(w-y)}dy$. Both functions $P_{\pm}$ are such that $0<P_{\pm}<1$ and $P_{+}(z)>P_{-}(z)$ so that $1/\left[1-\beta P_{+}(z)\right] > 1/\left[1-\beta P_{-}(z) \right]$ for $0<\beta<1$ (see Supplementary Appendix 2). Thus, the critical strain for delamination is always larger for the stored energy than the total work. This agrees with physical intuition, since in the case of the stored energy, only the elastic energy at any given time is available to drive fracture. However, in the case of total work, some of the dissipated energy may have gone into fracture. As compared to the empirical derivation in Equation \ref{TimeScales}, the solutions for critical time $t_c$ (or equivalently strain $\epsilon_c$) obtained from the energy balance approach involve more complex non-linear time-convolution functions $P_{\pm}$. Nevertheless, they again reveal the dependence of $t_c$ (or $\epsilon_c$) on three physical dimensionless quantities. This further confirms the important role of these dimensionless parameters in controlling the wrinkle-induced delamination process of the viscoelastic foulant layer. 

The wrinkling strain $\epsilon_w$ is assumed to be small, therefore the energy and relaxation in the foulant layer prior to wrinkling can be neglected in the above analysis. The effect of $\epsilon_w$, however, is taken as a shifting parameter to the solution obtained above as suggested in Pocivavsek et al. \cite{Luka2018}: $\epsilon_c-\epsilon_w$. Shown in Figure \ref{SchematicEApproach}b are the critical strains for delamination obtained from solving Equation \ref{Eqn4} when the three physical dimensionless parameters are varied. While a decrease in $E_\infty/E_0$ or in $\dot{\epsilon} \tau_R$ leads to an increase in the critical strain, a decrease in $\epsilon_{de}$ reduces the critical strain. However, as the figure shows, the effects are highly non-linear in the intermediate regime between the two elastic limits. In addition, smaller $E_\infty/E_0$ values correspond to larger differences between the solutions using total work $W$ and the stored energy $U$ in the energy balance approach. When there is no intrinsic material relaxation, $\left(E_\infty/E_0=1\right)$, the solutions from both approaches coincide to the prior linearly elastic case \cite{Luka2018}. Furthermore, in the two elastic limits $\dot{\epsilon}\tau_R \gg 1$ and $\dot{\epsilon}\tau_R \ll 1$, the two limiting elastic solutions $\epsilon_c-\epsilon_w = \epsilon_{de}$ and $\epsilon_c-\epsilon_w = (E_0/E_\infty)\epsilon_{de}$ are obtained. Analyzing the behavior of the analytical functions $P_{\pm}(z)$ in Equation \ref{Eqn4} confirms the same limiting behavior (see Supplementary Appendix 2). It is important to note that $\epsilon_c-\epsilon_w$ does not simply increase linearly with the dimensionless parameter $\epsilon_{de}$. Instead, we observe a right shift (in the arrow direction shown in Figure \ref{SchematicEApproach}b) for the transition from the instantaneous response to the other regimes which can be attributed to the non-linear effect of $\tau_{de}/\tau_R$.

\subsection{Finite Element Analysis} 
\begin{figure}
\centering
\includegraphics[width=1\linewidth]{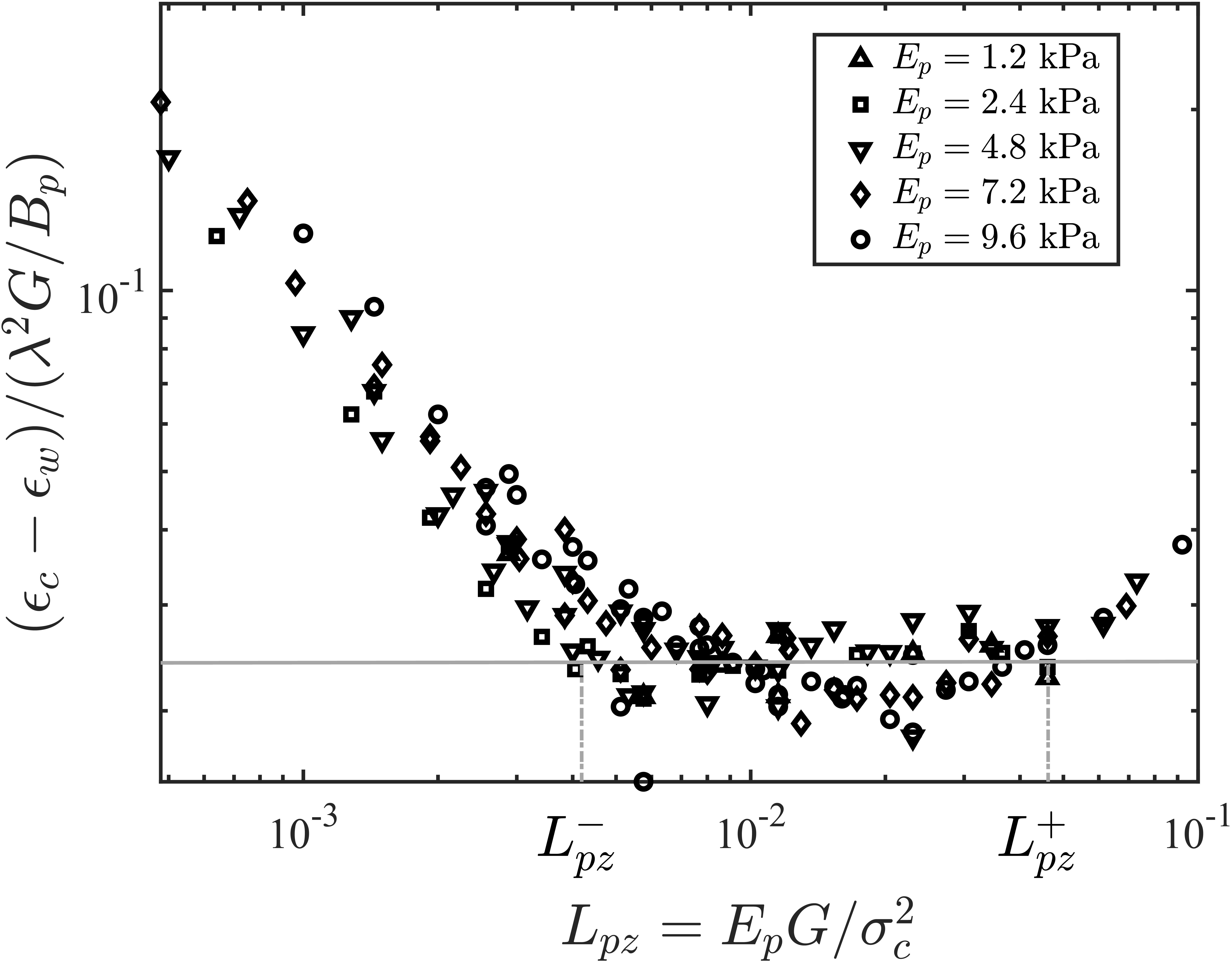}
\caption{Normalized ratio (or the slope/pre-factor $a$ in the scaling law: $\epsilon_c-\epsilon_w = a \lambda^2G/B_p$) versus $L_{pz}$. Various values of $E_p$ are examined. The flat line is a fit showing the linear scaling as predicted from scaling analysis which results in a slope of the same order as the one computed for the elastic foulant layer \cite{Luka2018} (see Supplementary Appendix 3).}
\label{CZMLengthEffect}
\end{figure}

\begin{figure*}
\centering
\includegraphics[width=1.0\linewidth]{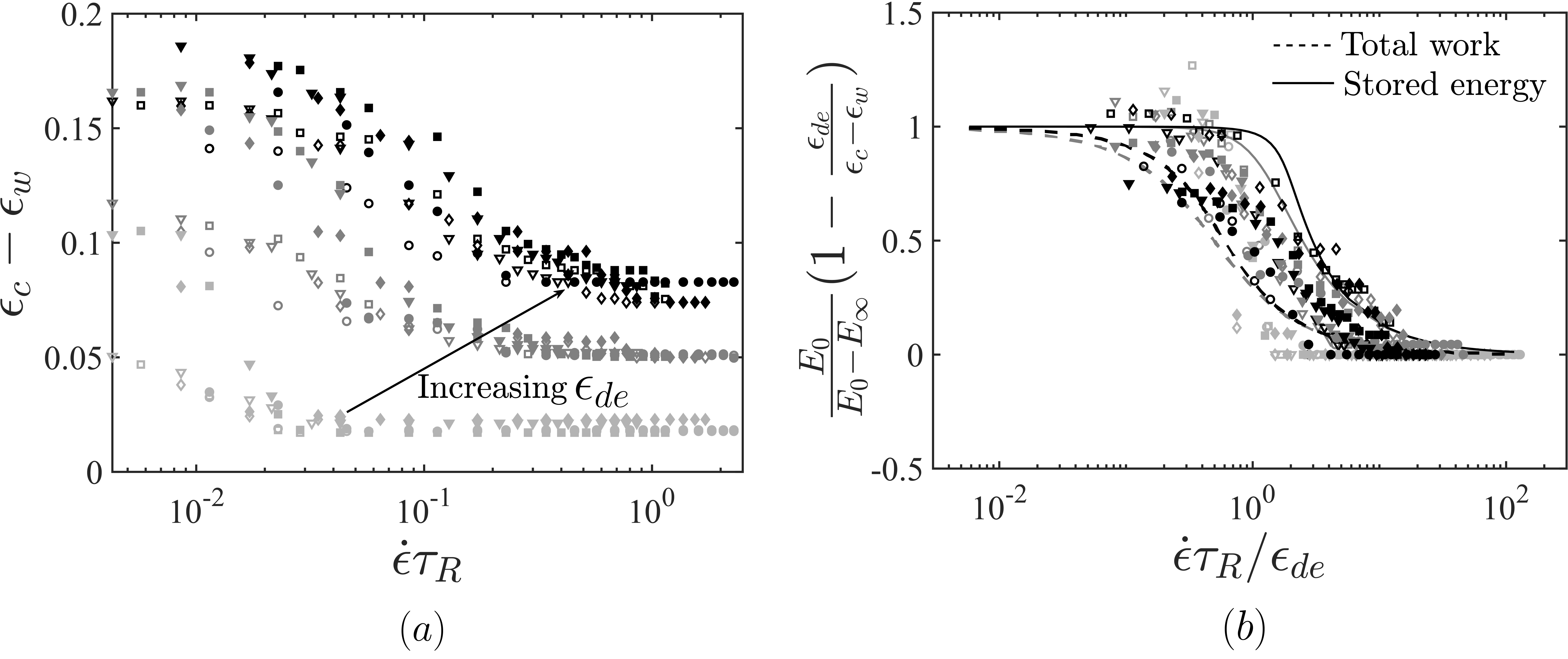}
\caption{Finite element results show how the three controlling dimensionless parameters affect the critical strain (a) and general fits to collapse numerical and analytical data (b). Four marker types (downward pointing triangle, square, diamond, circle) correspond to four loading rates $\dot{\epsilon}/\dot{\epsilon}_{el}=0.75, 1, 1.5, 2$, respectively used in FE simulations, where $\dot{\epsilon}_{el}$ is the loading rate adopted from the FE simulations in the previous study for elastic foulant layer \cite{Luka2018} (see Supplementary Appendix 3). Open and closed markers are for $E_\infty/E_0=0.5$ and $E_\infty/E_0=0.25$, respectively. Increasing levels of gray color of the markers for the numerical data correspond to increasing values of $\epsilon_{de}$; however, the color code and line styles in (b) for the analytical results are consistent with Figure \ref{SchematicEApproach}b.} 
\label{FEGeneralResult}
\end{figure*}

To further study the roles of the dimensionless control parameters and their influence on $\epsilon_c$, as well as verify the trends observed from the analytical method, a detailed parametric study is performed using finite element method (FEM) with cohesive zone model (CZM) implemented in Abaqus (Dassault Syst\`emes, MA) \cite{Abaqus18}. In this approach, traction separation laws are prescribed between the foulant layer and the film interface to study the delamination process \cite{Mei2011, Thouless2007,Turon2007, Xie2006, Heinrich2012, Paul2012, NN2016, Lin2019, NN2017, Thouless2019, Luka2018}. Two input parameters, the cohesive strength $\sigma_c$ and the fracture energy $G$, together with a damage law are necessary to describe the computational cohesive laws. In this study, bi-linear softening laws are employed to improve the numerical implementation in the previous study which employed linear elastic brittle laws \cite{Luka2018} (see Supplementary Appendix 3). The contributions of the two CZM parameters $\sigma_c$ and $G$ to the fracture process are also conveniently studied with this implementation. Solving equilibrium equations set by balancing the total energy, including contributions from both the foulant layer and the cohesive interface, allows the determination of $\epsilon_c$. In Abaqus, this solution process can be based on an implicit or explicit solver \cite{Abaqus18}. The implicit solver offers the advantage of solving this quasi-static problem without introducing additional dynamic energy. However, the presence and coupling of surface instability, material softening, contact conditions, and interfacial delamination in this problem requires tuning of various solver parameters and the introductions of certain artificial energies, such as damping, to resolve convergence issues of this iterative solution scheme. Therefore, in this study, we employ Abaqus dynamic explicit solver in order to conduct a parametric study with minimum adjustment of solver parameters over a large space of material properties and varying loading rates. Furthermore, in the CZM method, the process zone length along the interface $L_{pz} = EG/\sigma_c^2$ , which is the length over which CZM elements enter the degradation part of the traction separation law, affects the delamination mechanism in the problem. It has been shown in the literature for several classical interfacial geometries that $L_{pz}$ has to be smaller than a characteristic length of the system for CZM to produce the same solution as the energy based approach, such as $L_{pz}<h$ for double cantilever beam, where $h$ is the beam thickness \cite{Thouless2007, Heinrich2012}, or $L_{pz}<L^2/(2h)$ for edge delamination, where $L,h$ are the length and thickness of the layer \cite{Thouless2019}. The role of $L_{pz}$ was not considered in prior topography-driven delamination work \cite{Luka2018}. Thus, here we conducted a detailed sensitivity study for the effects of CZM parameters on $\epsilon_c$ in the instantaneous and long-term limits for the viscoelastic foulant layer. At these limits, the viscoelastic foulant layer can be treated as an elastic foulant layer with modulus $E_p=E_0=E_\infty$ and the scaling law becomes: $\epsilon_c =a \lambda^2G/B_p$, where $a$ is a constant pre-factor \cite{Luka2018}. Figure \ref{CZMLengthEffect} plots the normalized ratio $(\epsilon_c - \epsilon_w)/( \lambda^2G/B_p)$, which provides the parameter $a$, as a function of $L_{pz}$ for different foulant layer's modulus $E_p$. For each foulant layer's modulus, simulations with different sets of CZM parameters $\sigma_c$ and $G$ were performed. 
As shown in Figure \ref{CZMLengthEffect}, the presence of a process zone in CZM might influence the transition of different delamination mechanisms. We can compare the predictions from our analytical methods, which use only fracture toughness, with the FEM simulations only in the regime where the FEM data collapse to a flat line corresponding to a constant parameter for $a$. In this regime, fracture is dominated by energy, and the cohesive strength has negligible effect. Outside this regime, the strength might play a significant role and hence the scaling law $\epsilon_c =a \lambda^2G/B_p$ cannot be used to interpret the FEM data. As the scope of this paper is on the wrinkle-induced delamination mechanism of a viscoleastic foulant layer, focus on determination of the transition between these delamination mechanisms in FEM simulations will be presented in a separate publication. A summary is presented in Supplementary Appendix 3 to emphasize that the CZM analysis is conducted with careful attention to important numerical aspects including mesh refinement, process zone length, strength, and energy dominated regimes \cite{Thouless2007,Turon2007, Xie2006, Heinrich2012, Paul2012, NN2016, Lin2019, NN2017,Thouless2019}. Our simulation results for the delamination of a viscoelastic foulant layer from a wrinkled surface in the energy regime are plotted in Figure \ref{FEGeneralResult}a. As compared with Figure \ref{SchematicEApproach}b, the FE data confirm a similar dependence of the critical strain on the three dimensionless control parameters. A consistent right-shift of the transition from the instantaneous response to the other regimes as $\epsilon_{de}$ increases is also observed. 
\subsection{General Solution} Our analytical method and simulations indicate that the delamination of a viscoelastic foulant layer from a wrinkled surface shows a complex dependence on three control parameters. We therefore investigate the design parameter space by deriving a general fit to collapse the simulation data for a wide range of these parameters. Equation \ref{Eqn4} can be rewritten: 
\begin{equation}
\frac{\epsilon_c-\epsilon_w}{\epsilon_{de}} = \frac{1}{1-\beta P_{\pm}\left[(\epsilon_c-\epsilon_w))/(\dot{\epsilon}\tau_R)\right]}.
\label{Eqn5}
\end{equation}
Equation \ref{Eqn5} implies the dimensionless solution: $\frac{\epsilon_c-\epsilon_w}{\epsilon_{de}}=\Pi \left(\frac{\dot{\epsilon}\tau_R}{\epsilon_{de}},\beta\right)$. Note that the parameter $\dot{\epsilon}\tau_R/\epsilon_{de}$ is just the inverse of the parameter $\tau_{de}/\tau_R$ that is shown in Equation \ref{TimeScales} to control the instability. Similarly, Equation \ref{Eqn5} can be written as: $\frac{1}{\beta}\left(1-\frac{\epsilon_{de}}{\epsilon_c-\epsilon_w}\right)= P_{\pm}\left[(\epsilon_c-\epsilon_w)/(\dot{\epsilon}\tau_R)\right]$. Thus, now we can replace the general dimensionless solution with: 
\begin{equation}
\frac{1}{\beta}\left(1-\frac{\epsilon_{de}}{\epsilon_c-\epsilon_w}\right)= P_{\pm}\left[\Pi \left(\frac{\dot{\epsilon}\tau_R}{\epsilon_{de}},\beta\right)\right],
\label{Eqn6}
\end{equation}
which predicts the collapse of all the data into a region defined by the bounded functions $0<P_{\pm}<1$. With the use of this general solution, all data for the analytical method can be collapsed to two general curves corresponding to either the use of $W$ or $U$, as shown in Figure \ref{FEGeneralResult}b. FEM data is more noisy due to various numerical factors in the simulations such as the determination of the onset of delamination; nevertheless, they also nicely collapse into the small area bounded by the two analytical solutions. The FEM data also show that the amount of energy available to drive fracture in the simulations is bounded by the two limiting cases considered in the analytical approaches. The upperbound and lowerbound curves are physically intuitive as the stored energy provides the least amount of possible energy while the total work provides the largest possible energy available to drive fracture. As discussed above, we conducted FEM analysis using an explicit solver. We controlled the dynamic effect to be small to represent a quasi-static condition; however, certain amount of dynamic energy due to the solution process and dynamic propagation of the crack tip may still be present in these highly non-linear simulations, contributing to a difference in the amount of available energy in the simulations as compared to either limits in the analytical approach. Though neither of them exactly overlaps the numerical data from analytical methods, they provide good insight into the controlling factors in this system, and how to use them to control the delamination process. The general collapse obtained in Figure \ref{FEGeneralResult}b further confirms a reasonable agreement between the theoretical and numerical predictions in how the wrinkle-induced delamination mechanism of the viscoelastic foulant layer is controlled by different physical parameters related to the viscoelastic properties and loading process. Equation \ref{Eqn6} shows that the three physical parameters can further be combined into two dimensionless parameters, $\dot{\epsilon}\tau_R/\epsilon_{de}$ and $\beta$, significantly reducing the design parameter space needed for optimizing this complex system. 
\section{Discussion}
Topography-driven surface renewal is a powerful new mechanism to drive interfacial fracture over large surfaces at risk for fouling. The initial theory was limited to an elastic response regime for the fouling layer \cite{Luka2018}. Nevertheless, the mechanism shows promising applications in medical device design, specifically anti-thrombotic vascular grafts \cite{Luka2019, Nandan2020}. The application of actuation into medical devices requires precise knowledge of target strains, which will inform the choice of graft materials, sources of actuation loading, and range of biofoulants that the mechanism will effectively remove from the surface. In this paper, we substantially enrich the existing theory on actuating topography to account for both dynamics in the surface (strain rate $\dot{\epsilon}$) and the viscoelastic nature of the fouling layer. The results of this paper are directly applicable to the design of actuating biomedical devices targeting viscoelastic foulants such as thrombus, bacteria, and biofilms \cite{Luka2019, Nandan2020,Shivapooja2013,Levering2014}.

In summary, our FE results show consistent agreements with analytical predictions of the non-linear interplay between the three control parameters, which suggest several conditions to promote delamination in the viscoelastic regime. With the same rate of loading and adhesion energy, the critical strain for delamination onset decreases with increasing relaxation time of the foulant layer $\tau_R$. In other words, a foulant layer that relaxes slowly is easier to de-adhere. The limit of very fast relaxation, in which the thin foulant layer approaches the fluid-like limit, needs further investigation, as FE simulations in this regime indicate that another mechanism involving the critical strength, rather than only fracture toughness, may play a role here. With the same adhesion energy and same relaxation time, increasing loading rate $\dot{\epsilon}$ reduces the critical strain for the wrinkle-induced delamination onset. Furthermore, with the same relaxation time and rate of loading, decreasing adhesion energy also decreases the critical strain for the wrinkle-induced delamination onset. In addition, comparisons are also made for the energy balance approach using the stored energy and the total energy, and a general collapse of simulations data to a bounded regime between these analytical predictions help to reduce the design parameter space. Thus, this work provides a first attempt towards understanding the delamination of realistic viscoelastic biofoulants.

Ultimately, topographic surface renewal is based on an energy release mechanism whereby external loading and deformation of an adherent foulant leads to accumulated energy in the foulant, which beyond a critical value is able to drive interfacial crack propagation. In the purely elastic case, as shown in the key result of our prior work (see Equation 2 in Pocivavsek et al. \cite{Luka2018}), the critical strain density needed to drive surface renewal is given by $\epsilon_{c}/\lambda^2 \sim 1/\ell_{ec}^2$, where $\epsilon_{c}$ is the critical nominal compressive strain in the bilayer substrate driving surface wrinkling and foulant deformation and $\ell_{ec}=(B_0/G)^{1/2}$ is the elasto-capillary length scale. Because energy dissipation only occurs with interfacial fracture, the elastic case is independent of loading history; as such it is independent of actuation in the dynamic sense. 

Unlike in the elastic case, the presence of viscoelasticity in the foulant introduces a second intrinsic mode of dissipation in addition to interfacial fracture. Furthermore, the strain state of a viscoelastic foulant layer is a superposition of states because of the intrinsic material softening triggered by loading. The coupling of the non-linear, wrinkle-induced strain field with the non-linear dependence of the strain state on loading leads to a highly non-linear equation to determine the critical delamination strain for a viscoelastic foulant layer. We show that this critical strain is controlled by three physical quantities: the magnitude of foulant layer relaxation $E_\infty/E_0$, the rate of material relaxation relative to loading rate $\dot{\epsilon}\tau_R$, and the critical strain to delaminate the foulant layer in the instantaneous elastic limit $\epsilon_{de} \sim\lambda^2G/B_0$. A fourth dimensionless parameter is the strain to initiate wrinkling $\epsilon_w$, which can be tuned by graft construction and made negligibly small in the case of vascular graft design \cite{Genzer2006,Pocivavsek2009,Luka2019, Nandan2020}. In the first part of this paper, we show that viscoelasticity, characterized as the decay of foulant stiffness from $E_0$ to $E_\infty$ over a characteristic time scale $\tau_R$, breaks the critical strain for topographic de-adhesion into two limiting regimes: $\epsilon_{de} \leq \epsilon_c -\epsilon_w \leq \left(E_0/E_\infty\right) \epsilon_{de}$. The smallest strain is set by the elastic limit \cite{Luka2018}. However, the critical strain increases proportionally to the degree of degradation in stiffness to an upper bound set by $E_0/E_\infty$.  This inequality can be directly obtained from our general solution by solving for $\epsilon_c$ in Equation \ref{Eqn6} or equivalently by using Figure \ref{FEGeneralResult}b to determine the value of $f=P_{\pm}\left[ \Pi \left( \frac{\dot{\epsilon}\tau_R}{\epsilon_{de}},\beta\right)\right]$
\begin{equation}
\epsilon_c= \epsilon_{de} \left[1- f\left(1-\frac{E_\infty}{E_0} \right) \right]^{-1} + \epsilon_w,
\label{Eqn7}
\end{equation}
where $f \in [0,1]$ is the range of the vertical axis in Figure \ref{FEGeneralResult}b. By moving $f$, the critical strain moves between the short and long-time limits which are connected by a highly non-linear transition region. Equation \ref{Eqn7}, and equivalently Figure \ref{FEGeneralResult}b, provide a unified approach to optimize strategies for actuated topographic surfaces designed to remove viscoelastic foulants. The horizontal axis in Figure \ref{FEGeneralResult}b is controlled by $\frac{\dot{\epsilon}\tau_R}{\epsilon_{de}} = \left( \frac{\epsilon/\lambda^2}{\tau} \right) \cdot \left(\tau_R \ell_{ec}^2 \right)$, where $\tau$ is the applied actuation time. The first set of parentheses contains parameters under direct control of the graft designer, and the second set of parentheses incorporates intrinsic properties of the foulant. 

An effective anti-fouling design must enforce the system to work under short time conditions to lower the critical strain to the value $\epsilon_{de}$ which is achieved when $f=0$. Physically, this is equivalent to a state where intrinsic viscoelastic dissipation has not had time to take effect and all accumulated strain energy is available to drive interfacial fracture. Following the results in Figure \ref{FEGeneralResult}b and Equation \ref{Eqn6}, a sufficient condition to be in this regime is to satisfy $\frac{\dot{\epsilon}\tau_R}{\epsilon_{de}} \gtrsim10$. Thus, a system designed to be strained to a value $\epsilon$ along a time scale $\tau$ must satisfy the condition $\epsilon/\tau \gtrsim10 \epsilon_{de}/\tau_R$. Equivalently, the system must be designed to have a strain rate larger than $\epsilon_{de}/\tau_R$. In terms of the elasto-capillary length scale whose value depends on the specific response of the foulant, we obtain the following optimal design condition using Figure \ref{FEGeneralResult}b:
\begin{equation}
\frac{\epsilon/\lambda^2}{\tau} \gtrsim10 \frac{1/l_{ec}^2}{\tau_R}.
\label{Eqn8}
\end{equation}
Equation \ref{Eqn8} provides the condition for optimal surface renewal for a given viscoelastic foulant.

In general, if one moves $f$ up from 0, then the critical strain given by Equation \ref{Eqn7} increases. This is important from a design perspective as it provides a design criterion to improve the de-adhesion capability of the system if the optimal surface renewal condition cannot be achieved. For instance, if the system is forced to operate at $\frac{\dot{\epsilon}\tau_R}{\epsilon_{de}}\gtrsim5$ then the largest value of $f$ becomes 0.3 and the critical strain needed will be given by Equation \ref{Eqn7}. 

In topography-driven surface renewal for viscoelastic foulants, the fouling layer accumulates strain density ($\epsilon/\lambda^2$) over an applied actuation time ($\tau$). We will take the example of vascular grafts as an illustration, although similar arguments can be constructed for any specific application. The designer has full control over all three parameters independently: $\epsilon$ is the nominal strain in the graft substrate and will be set by $E_s$ and the load available in the system (for example, pulse pressure in vascular grafts \cite{Luka2019,Nandan2020}), surface wavelength $\lambda$ is tunable via graft bilayer construction \cite{Genzer2006,Pocivavsek2009}, and lastly $\tau$ is either set by the system (heart rate in case of vascular grafts \cite{Luka2019,Nandan2020}) or, if externally driven, by an actuating power source (such as in soft robotic on-demand fouling-release urinary catheters \cite{Shivapooja2013,Levering2014}). These three parameters must combine such that the time-integrated accumulated strain density is within the limits set by purely intrinsic properties of the foulant layer: $G$, $E_0$, $E_\infty$, $h$, and $\tau_R$. Our analysis allows a concrete, quickly applicable methodology to pick design parameters for given anti-fouling applications of topography-driven surface renewal. It shows that to optimize topographic de-adhesion for viscoelastic foulants the condition set forth in Equation \ref{Eqn8} should be satisfied. This helps guide design of anti-fouling surfaces targeted for biomedical applications such as anti-thrombotic grafts.    

\begin{acknowledgments}
E.C. and E.H. acknowledge the support of Fondecyt Grant No 1201250. S.V. acknowledges the support of NSF-CMMI 1824708. L.P. and N.N. acknowledge the support of the grant NIH-1R01HL159205-01. We thank the Center for Research Informatics (CRI) which is funded by the Biological Sciences Division at the University of Chicago with additional funding provided by the Institute for Translational Medicine, CTSA grant number UL1 TR000430 from the National Institutes of Health.
\end{acknowledgments}

\section*{Data Availability Statement}
The data that support the findings of this study are available from the corresponding author upon reasonable request.

\appendix

\section*{Appendix 1: Mechanical analog and energy components for a viscoelastic solid}
Consider a viscoelastic Maxwell solid shown in Figure 6. By considering either a displacement controlled experiment or a force controlled experiment applied to this system, it is straightforward to see that this system yields the right behavior as the viscoelastic model with a single Prony series: $E(t)=E_\infty+(E_0-E_\infty)e^{-t/\tau_R}$ used in the main text. In order to physically understand different energy components in a viscoelastic system, a mass is added to the end of this system to include inertial effects. By applying a force $F$ to the mass, the following equation of motion is  obtained: 
\begin{eqnarray}
m \ddot{x}&=&-E_{\infty}x-(E_0-E_\infty)x_1+F,
\label{Maxwell}
\end{eqnarray}
where $x$ is the total displacement of the system and $x_1$ is the displacement of the spring with stiffness $E_0-E_\infty$. The dashpot is always in equilibrium with the contiguous spring and follows the equilibrium equation
\begin{eqnarray}
\eta \dot{x_2}&=&(E_0-E_\infty)x_1,
\end{eqnarray}
where $x_2=x-x_1$ corresponds to the dashpot displacement. It defines the relaxation time $\tau_R=\eta/(E_0-E_\infty)$ that accounts for different response regimes of the system. For quasi-static motion or zero inertial effects ($m \ddot{x}=0$), the applied force is in balance with the springs and dashpot $F=E_{\infty}x+(E_0-E_\infty)x_1$. For short times, $t\ll \tau_R$, the dashpot does not have sufficient time to react and $x_2\approx 0$. It yields the force-displacement relation $F=E_0 x$. For large times, $t\gg \tau_R$, the dashpot has time to relax and $x_2\approx x$. It means that the spring with stiffness $E_\infty$ takes the total load applied to the system and $F=E_\infty x$. In general, the force-displacement relation is 
\begin{eqnarray}
F(t)=E_0 x(t)+\int_0^t d\tau x(\tau) \frac{d}{dt}E(t-\tau),
\end{eqnarray}
where $E(t)=E_\infty+(E_0-E_\infty)e^{-t/\tau_R}$.

The energy balance is obtained by multiplying Equation \ref{Maxwell} by $\dot{x}$ and integrating by parts:
\begin{eqnarray}
\frac{d}{dt}\left[\frac{m}{2}\dot{x}^2+\frac{E_\infty}{2}x^2+\frac{E_0-E_\infty}{2}x_1^2\right]=-\eta\dot{x_2}^2+F\dot{x}.
\end{eqnarray}
Here, we recognize the kinetic energy $T$,  the stored energy $U$, and the total energy $E=T+U$, which can be defined as: 
\begin{eqnarray}
E&=&T+U,\nonumber\\
T&=&\frac{m}{2}\dot{x}^2,\nonumber\\
U&=&\frac{E_\infty}{2}\dot{x}^2+\frac{E_0-E_\infty}{2}\dot{x_1}^2.
\end{eqnarray}
This leads to
\begin{eqnarray}
\frac{dE}{dt}=-\eta \dot{x_2}^2+F\dot{x},
\end{eqnarray}
from which the rate of dissipation can be recognized as $\dot{D}=\eta \dot{x_2}^2$ and the external work per unit of time $\dot{W}=F\dot{x}$. In other words, 
\begin{eqnarray}
\frac{dE}{dt}=-\dot{D}+\dot{W}.
\end{eqnarray}
This states that the total mechanical energy decreases by dissipation and increases by the external work applied to the system\cite{Christensen2003}. Thus, by using the mechanical analog above, a physical view of different energy components is elucidated and the stored energy is the elastic energy of the two springs. The analysis can also be generalized to a continuum system. Consider the case of uniaxial compression, traction, or bending of a filament where we expect the following constitutive relation: 
\begin{eqnarray}
\sigma=E_{\infty}\epsilon+(E_0-E_\infty)\epsilon^e.
\end{eqnarray}
Here $\epsilon^e$ is not the complete strain $\epsilon$ because part of the strain is taken by a viscous term $\epsilon^v$ such that $\epsilon=\epsilon^e+\epsilon^v$. In this regard, $\epsilon^e$ and $\epsilon^v$ are equivalent to  $x_1$ and $x_2$, respectively,  in the previous model. With no viscous term, the effective stiffness of the system is $E_0$, however, the viscous term dissipates the elasticity of the contiguous spring with stiffness $E_0-E_\infty$ and the effective stiffness for long times is $E_\infty$. The stress of the spring with stiffness $E_0-E_\infty$ is in balance with a viscous stress: $\sigma^e=\sigma^v$. It yields: 
\begin{eqnarray}
(E_0-E_\infty)\epsilon^e&=&2\eta \dot{\epsilon}^v,
\end{eqnarray}
where we use the constitutive relation $\sigma^v=2\eta\dot{\epsilon}$ to describe the uniaxial deformation of a viscous fluid. Thus, the following representation for the viscous strain in terms of the total strain is obtained: 
\begin{eqnarray}
2\eta\dot{\epsilon}^v=(E_0-E_{\infty})(\epsilon-\epsilon^v).
\end{eqnarray}
Defining the relaxation time $\tau_R=2\eta/(E_0-E_{\infty})$, we obtain the equation: 
\begin{eqnarray}
\tau_R\dot{\epsilon}^v=\epsilon-\epsilon^v,
\end{eqnarray}
that allows us to find the viscous strain: 
\begin{eqnarray}
\epsilon^v=\frac{1}{\tau_R}\int_0^t \epsilon(\tau)e^{-(t-\tau)/\tau_R}d\tau.
\end{eqnarray}
Thus, the stored energy as the elastic energy of the two springs is generalized as: 
\begin{eqnarray}
U=\frac{E_\infty}{2}\int_V dV \epsilon^2  + \frac{E_0-E_\infty}{2} \int_V dV \left({\epsilon^e}\right)^2.
\end{eqnarray}
The above relation corresponds to the one presented by Christensen\cite{Christensen2003} where the stored energy is computed as: 
\begin{eqnarray}
U=\frac{1}{2}\int_{-\infty}^{t}\int_{-\infty}^{t} E(2t-\tau-\zeta)\frac{d\epsilon}{d\tau}\frac{d\epsilon}{d\zeta}d\epsilon d\zeta.
\end{eqnarray}
The total energy of the viscoelastic solid corresponds to the total injected work to the system, which is:
\begin{eqnarray}
\dot{W} = \int_V \sigma \dot{\epsilon} dV.
\end{eqnarray}  
\section*{Appendix 2: Analytical method for wrinkle induced delamination}
From the imposed wrinkled topography $h(x,t)=A(t)\sin(kx)$, the imposed strain in the foulant layer is: 
\begin{eqnarray}
\gamma=-yk^2A(t)\sin(kx),
\end{eqnarray}
where $A(t)=\frac{2}{k}\sqrt{\dot{\epsilon}(t-t_w)}$. Neglecting the part of the strain previous to buckling and taking $t_w=0$, the viscous strain is computed as: 
\begin{eqnarray}
\gamma^v=yk^2\frac{2}{k}\sin(kx)\sqrt{\dot{\epsilon}\tau_R}F(t/\tau_R),
\end{eqnarray}
where $F(x)=\int_0^x \sqrt{y}e^{-(x-y)}dy$. Therefore, the elastic strain is:
\begin{eqnarray}
\gamma^e=yk^2\sin(kx)\left[A(t)-A(\tau_R)F(t/\tau_R)\right].
\end{eqnarray}
Here $A(t)=\frac{2}{k}\sqrt{\dot{\epsilon}t}$ because of the approximations. 
We can now compute the stored energy and the total work. The stored energy should be the elastic energy of the two springs (Equation 21, Appendix 1) which becomes: 
\begin{eqnarray}
U/L&=&\frac{E_\infty I}{4}\left[k^2A(t)\right]^2\nonumber\\
&+&\frac{(E_0-E_\infty)I}{4}\left[k^2(A(t)-A(\tau_R)F(t/\tau_R))\right]^2.
\end{eqnarray}
The total work in Equation 23 of Appendix 1 becomes: 
\begin{eqnarray}
W/L=\frac{E_0 I}{4}\left[k^2A(t)\right]^2 - \frac{(E_0-E_\infty)I}{4}\left[k^2A(\tau_R)\right]^2 H(\tau/\tau_R),
\end{eqnarray}
where $H(z)$ is a dimensionless integral, so that 
\begin{eqnarray}
H(z)=\int_0^z dw \frac{F(w)}{\sqrt{w}}=zP_{-}(z),\\
P_{-}(z)=1 - {}_2F_2[1,1;\frac{3}{2},2;-z].
\end{eqnarray}
Note that ${}_2F_2[1,1;\frac{3}{2},2;-z]$ is a hypergeometric function of order $(2,2)$.

To drive the fracture in a viscoelastic system, approaches balancing either the total work or the stored energy against the fracture toughness have been proposed in the literature\cite{Roger2013,Ravichandran1994}. We first start with the consideration of the balance using the total work: $0=d(W+U_S)$, where the surface energy is $U_S=Gbl$, where $G$ is the work of fracture, $b$ is the film width, $l$ is the crack length. This leads to: 
\begin{eqnarray}
\frac{B_0}{4}\left[k^2A(t)\right]^2-\frac{B_0-B_\infty}{4}\left[k^2A(\tau_R)\right]^2H(t/\tau_R)=G.
\end{eqnarray}
Using the relation $H(z)=zP_{-}(z)$, this can be rewritten: 
\begin{eqnarray}
\frac{B_0}{4}\left[k^2A(t)\right]^2-\frac{B_0-B_\infty}{4}\left[k^2A(t)\right]^2P_{-}(t/\tau_R)=G.
\end{eqnarray}
Therefore, 
\begin{eqnarray}
t_c/\tau_R \sim \frac{\lambda^2 G/(B_0\dot{\epsilon}\tau_R)}{\left[1-\beta P_{-}(t_c/\tau_R)\right]}.
\end{eqnarray}
For the energy balance approach using the stored energy: $0=d(U+U_S)$, the following relation is obtained: 
\begin{eqnarray}
\frac{B_\infty}{4}\left[k^2A(t)\right]^2+\frac{B_0-B_\infty}{4}\left[k^2\left(A(t)-A(\tau_R)F(t/\tau_R)\right)\right]^2=G.
\end{eqnarray}
Defining the function $P_{+}(z)=\left[1-\left(1-\frac{F(z)}{\sqrt{z}}\right)^2\right]$, this relation is rewritten as follows: 
\begin{eqnarray}
\frac{B_0}{4}\left[k^2A(t)\right]^2-\frac{B_0-B_\infty}{4}\left[k^2A(t)\right]^2P_{+}(t/\tau_R)=G.
\end{eqnarray}
Therefore, 
\begin{eqnarray}
t_c/\tau_R \sim \frac{\lambda^2 G/(B_0\dot{\epsilon}\tau_R)}{\left[1-\beta P_{+}(t_c/\tau_R)\right]}.
\end{eqnarray}
Thus, the non-linear equations to determine the critical delamination time for both approaches have the same following forms: 
\begin{eqnarray}
t_c/\tau_R \sim \frac{\lambda^2 G/(B_0\dot{\epsilon}\tau_R)}{\left[1-\beta P_{\pm}(t_c/\tau_R)\right]},
\end{eqnarray}
with $P_{+}$ and $P_{-}$ defined above. Note that these two functions are bounded between 0 and 1, and $P_{+}(z)>P_{-}(z)$. We observe the followings: \\
For $\lambda^2 G/(B_0\dot{\epsilon}\tau_R) \ll1$ and $\beta\ll1$, then 
\begin{eqnarray}
t_c/\tau_R \sim {\lambda^2 G/(B_0\dot{\epsilon}\tau_R)}.
\end{eqnarray}
For $\lambda^2 G/(B_0\dot{\epsilon}\tau_R) \gg1$ , then 
\begin{eqnarray}
t_c/\tau_R \sim \frac{\lambda^2 G/(B_0\dot{\epsilon}\tau_R)}{1-\beta}.
\end{eqnarray}
The above non-linear equations are solved in MATLAB (Mathworks, MA) for varying combinations of dimensionless parameters and results are shown in Figure S3b of the main text. 
\section*{Appendix 3: Numerical modeling with finite element (FE) and cohesive zone method (CZM)}
Details of the FE  model using Abaqus Explicit for the tri-layer system (foulant layer, film, and substrate) with the specified boundary conditions in the main manuscript are provided in the previous study\cite{Luka2018}. Each layer is modeled with three-dimensional solid elements (C3D8R, 8-node linear brick, reduced integration, and with hourglass control). The front and back faces are constrained in the z direction to maintain an effective plane strain condition. The foulant layer is described by the linear viscoelastic constitutive relationship with a single Prony series: $E(t)=E_\infty+(E_0-E_\infty)e^{-t/\tau_R}$ while the film and the substrate are modeled as incompressible, neo-Hookean materials. The modulus of the foulant layer $E_p(t)$ is much smaller than the modulus of the film $E_f$ such that $B_p \ll B_f$. The ratio between the moduli of the film and the substrate is $E_f/E_s \sim 80$. In order to model delamination of the foulant layer, the interface between the foulant layer and the film is modeled using Abaqus cohesive interface\cite{Abaqus18} with bi-linear traction separation laws as shown in Figure 7. This bi-linear shape for the traction separation law reduced numerical singularity caused by the abrupt drop of stress in the right-triangular traction profile used in the previous study\cite{Luka2018}. 

Though mixed-mode delamination\cite{Hutchinson1992,NN2017} can be important along this wrinkled interface, to simplify the problem, here independent traction separation laws as shown in Figure 7 for the normal and tangential modes are used in this study. In addition, due to the lack of experimental data for the critical strength $\sigma_c, \tau_c$ and fracture toughness $G_{IC},G_{IIC}$ for the biologically relevant interfaces considered here, $\sigma_c=\tau_c$ and $G_{IC}=G_{IIC}=G$ are assumed for the traction separation laws. This simplifying assumption may not be realistic as $G_{IIC}$ is often larger than $G_{IC}$, however, for computation simplification, such difference is neglected here\cite{NN2017}. Note that delamination only occurs when at least one CZM element on the interface reaches the end of the traction separation law. In other words, the stress must reach the critical value and the energy must be dissipated by an amount equal to $G$ so that the fracture surface can be created. Hence, both stress and energy are used as criteria for fracture onset as well as propagation in CZM. The delamination onset is followed by checking the CSDMG parameter in Abaqus, which reaches the value of 1 when the traction reaches the end of the traction separation law. The delamination of interest here is the one in which the foulant layer de-adheres without arrest as analyzed in the analytical model; hence, the point of delamination is also determined through checking the contact area between the foulant layer and the film. Prior to delamination, the area remains almost flat. When the critical point is reached, the contact area drops rapidly signifying an unstable detachment at their interface. 

Figure 8 shows typical stages in the compression process. When the nominal applied strain is smaller than the critical value for wrinkling to occur $\epsilon < \epsilon_w=\frac{1}{4} (3 E_s/E_f)^{2/3}=0.028$, the system remains flat (Figure 8a)\cite{Genzer2006, Luka2018, Allen1969, Bowden1998, Pocivavsek2008, Sun2012, Cao2012, Cerda2003}. Upon further compression above the critical value $\epsilon_w$, the bilayer wrinkles. The foulant layer conformally follows this wrinkled topography (Figure8b). When a critical amplitude $A_c$ is reached, the foulant layer starts detaching from the bilayer surface (Figure 8c). 

Utilizing this FE model with the prescribed CZM traction separation laws, we investigate the effect of $\sigma_c$ and $G$ on the onset of delamination. The mesh size is chosen such that at least 3-5 elements are inside the process zone length\cite{Turon2007} $L_{pz}=E_pG/\sigma_{c}^2$. Note that for the right-triangular traction separation law used in the previous study for the elastic foulant layer\cite{Luka2018}, $G=0.5 \sigma_c^2/K$, this length becomes $L_{pz} = E_p/K$, where $K$ is the initial stiffness of the traction law. Thus, the three parameters $\sigma_c$, $K$, and $G$ are correlated in this implementation, making it difficult to separate the influence of individual parameters. However, the bi-linear profile offers an advantage to address this limitation. The initial stiffness $K$ is set to a high value as required in CZM modeling and the influence of $\sigma_c$ and $G$ can be examined. Mesh sensitivity studies are conducted to show that similar results are obtained when the mesh size is refined. Furthermore, the effects of $\sigma_c$ and $G$  are also studied for several values of the foulant layer stiffness in order to connect the CZM approach and our analytical model. The following sets of material and geometric parameters are utilized for both solution approaches. The substrate thickness $h_s$ is much bigger than the film thickness $h_f$ so that the wrinkle pattern can be described as in the previous wrinkling work for bilayers\cite{Genzer2006,Luka2018,Allen1969,Bowden1998,Pocivavsek2008,Sun2012,Cao2012,Cerda2003}. The length of the system $L$ is chosen so that it covers at least 8-10 wavelengths $\lambda$. The foulant layer thickness is selected to be in the thin layer limit\cite{Luka2018} $h_p/\lambda <1$ , specifically here a ratio of $h_p/\lambda=0.25$ is used unless otherwise stated. The materials properties for the substrate and film, and the instantaneous modulus $E_p(0)$ of the foulant layer are adopted from the previous study\cite{Luka2018}. The relaxation time $\tau_R$ and the amount of relaxation $E_0/E_\infty$ are varied to study their influence on $\epsilon_c$. The compression is applied as described in the work of Pocivavasek et al.\cite{Luka2018}. Specifically, for the elastic case, a displacement velocity that smoothly increased over 0.5 ms to the target $v_x = 0.01$ mm/ms was prescribed to the two ends of the tri-layer system with a loading rate $\dot{\epsilon}_{el}=0.06$ ms$^{-1}$. 

At the instantaneous and long-term response, the viscoelastic foulant layer can be treated as an elastic foulant layer with modulus $E_p$ and the scaling law  becomes\cite{Luka2018}: $\epsilon_c-\epsilon_w = a \lambda^2G/B_p$. Using this scaling law, the normalized parameter $\widetilde{G}=\lambda^2G/B_p$ will be utilized to analyze the FEM results. FE simulations of the delamination onset for the two cases, $\sigma_c/E_p=0.15$ and $\sigma_c/E_p=0.25$, with $E_p=4.8$ kPa are shown in Figure 9. For each case, three states corresponding to $\Tilde{G} = 0.9, 1.8, 4.8$ are presented. As shown in the top three figures (Figure 9-a,b,c), at $\sigma_c/E_p=0.15$, the first two cases (a,b) of $\widetilde{G}$ have almost the same critical delamination strain $\epsilon_{c}-\epsilon_w \sim 0.04, 0.05$, indicating that $\widetilde{G}$ does not play a significant role here. However, when $\Tilde{G}$ increases, it starts to take effect, i.e. $\epsilon_{c}-\epsilon_w \sim 0.12$ for the third case (Figure9-c). The same trend is observed at $\sigma_c/E_p=0.25$ where $\epsilon_{c}-\epsilon_w \sim 0.08, 0.08, 0.1$ in Figure 9-d,e,f, respectively. Furthermore, for the third value of $\widetilde{G}$, the two cases (Figure 9-c,f) have similar $\epsilon_c$ indicating that the change in the strength from $\sigma_c/E_p=0.15$ to $\sigma_c/E_p=0.25$ does not significantly influence $\epsilon_c$.

These observations are consistent with the capacity of CZM to bridge different failure mechanisms as discussed in\cite{Thouless2019,Thouless2007,Heinrich2012}, but still requires further investigation in the context of our topography-driven delamination. As noted, the CZM process zone length, $L_{pz}$ plays a key role in determining the transitions between these mechanisms\cite{Thouless2019,Thouless2007,Heinrich2012}. Figure 9 in the main manuscript plots the ratio $(\epsilon_c-\epsilon_w)/\widetilde{G}$ with respect to $L_{pz} = E_pG/\sigma_c^2$. The flat region is interpreted as the constant value for the slope of the linear scaling law between $\epsilon_c-\epsilon_w$ and $\widetilde{G}$. It indicates that $\sigma_c$ is insensitive here and $\epsilon_c-\epsilon_w$ scales linearly with $\widetilde{G}_C$. However, if $L_{pz}$ is too small or too large, $\sigma_c$ might influence the results and deviate the solution from the energy based approach. Note that a pre-factor of approximately 0.025 is obtained from Figure 4 of the main manuscript. In order to compare this value with the one presented in the previous study for the case of elastic foulant layer \cite{Luka2018}, the correlation between amplitude and strain $A_c \sim \frac{\lambda}{\pi}\sqrt{\epsilon_c-\epsilon_w}$ can be used to determine the prefactor for the critical amplitude. This leads to a scaling: $A_c \sim 0.05 \lambda^2 (G/B_p)^{1/2}$ which is of the same order with the value $c=0.13$ presented previously\cite{Luka2018}. They are not identical because, as discussed above, the right-triangular traction separation law has several numerical disadvantages and the effects of individual CZM parameters have not been investigated for this case as compared to the use of the bi-linear laws. However, the scaling dependence on $\widetilde{G}$ is consistent with analytical analysis with the same order of the numerical pre-factor for both cases of traction separation laws. In order to illustrate this region, Figure 10 plots the results from FEM for different cases of foulant layer's modulus and thickness. For each case, only the results of the sets of CZM parameters where $\sigma_c$ has insignificant effect on $\epsilon_c$ are selected in this plot. The linear relationship observed in Figure 10 suggests that $\epsilon_c-\epsilon_w$ varies linearly with $\widetilde{G}$ as predicted by the scaling law. 

Taking into account these considerations for CZM, we conducted FEM simulations with the viscoelastic foulant layer for $L_{pz}$ in the energy dominated region $L_{pz}^{-}<L_{pz}<L_{pz}^{+}$. This allows a consistent comparison between FEM results and analytical model which focus on fracture driven by an energy release mechanisms. Results are presented in Figure 10 of the main manuscript.

\begin{figure*}
\centering
\includegraphics[width=0.5\textwidth]{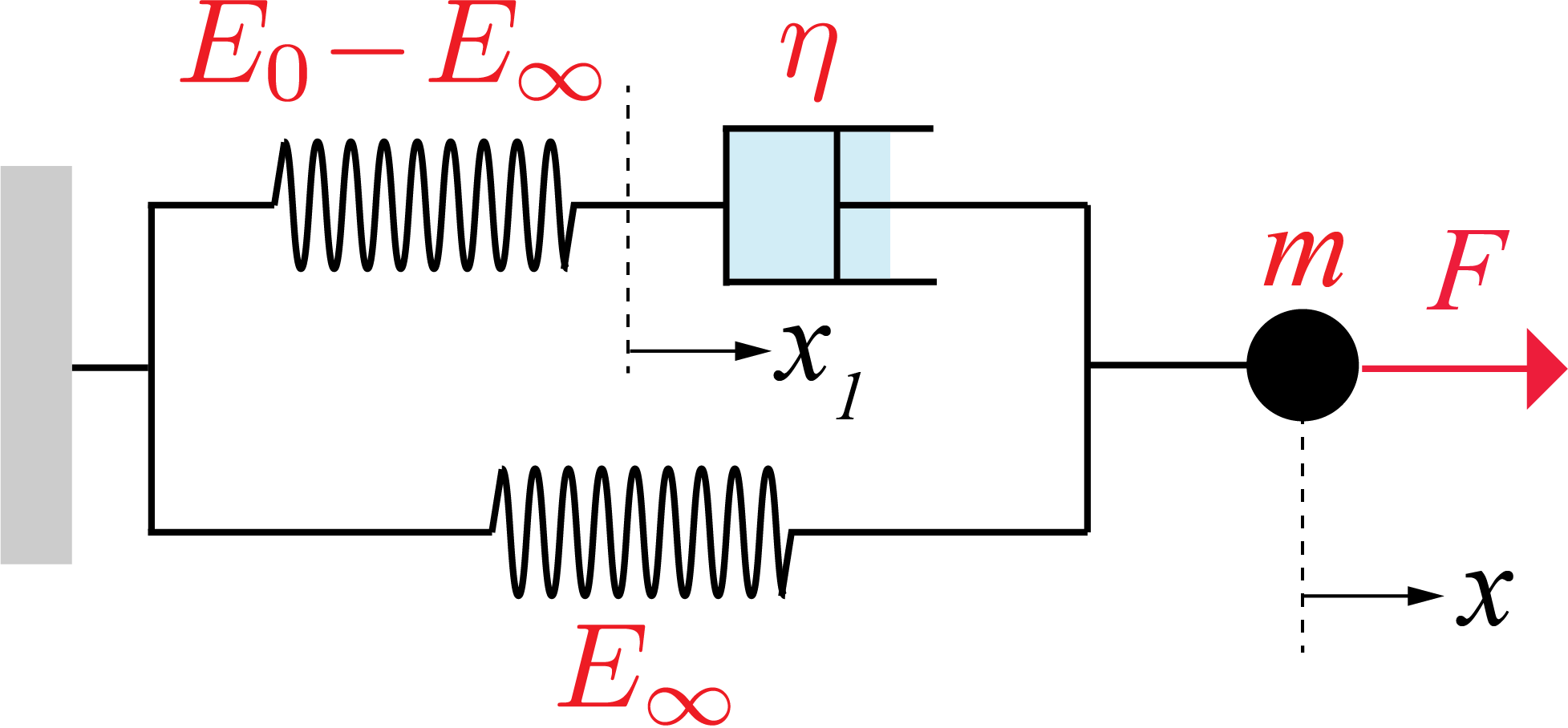}
\caption{A mechanical analog for a viscoelastic solid.}
\label{FigS1}
\end{figure*}
\begin{figure*}
\centering
\includegraphics[width=0.7\textwidth]{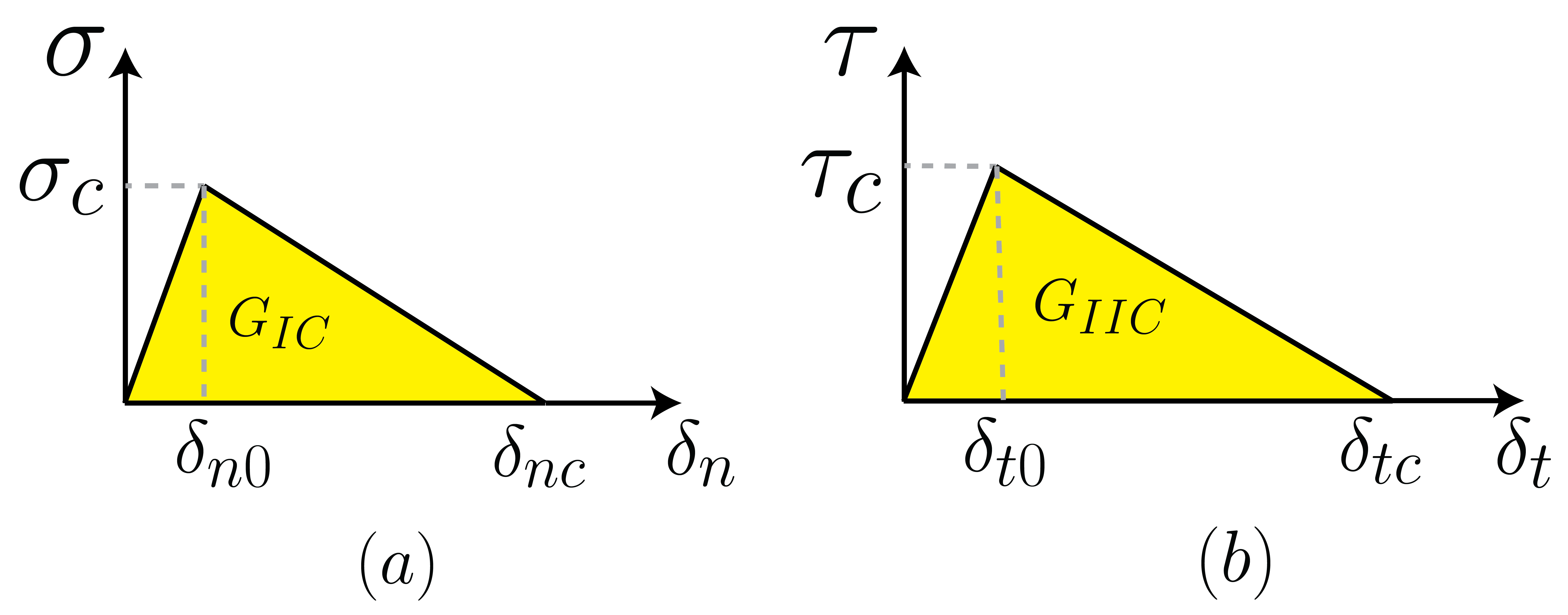}
\caption{Bi-linear traction separation laws for the normal and tangential modes. Delamination occurs when the tractions reach the ends of these traction laws and the separations attain the critical values $\delta_{nc}$ and $\delta_{tc}$ at which the energies $G_{IC}$ and $G_{IIC}$ are released to create the new surface.}
\label{FigS2}
\end{figure*}
\begin{figure*}
\centering
\includegraphics[width=0.7\textwidth]{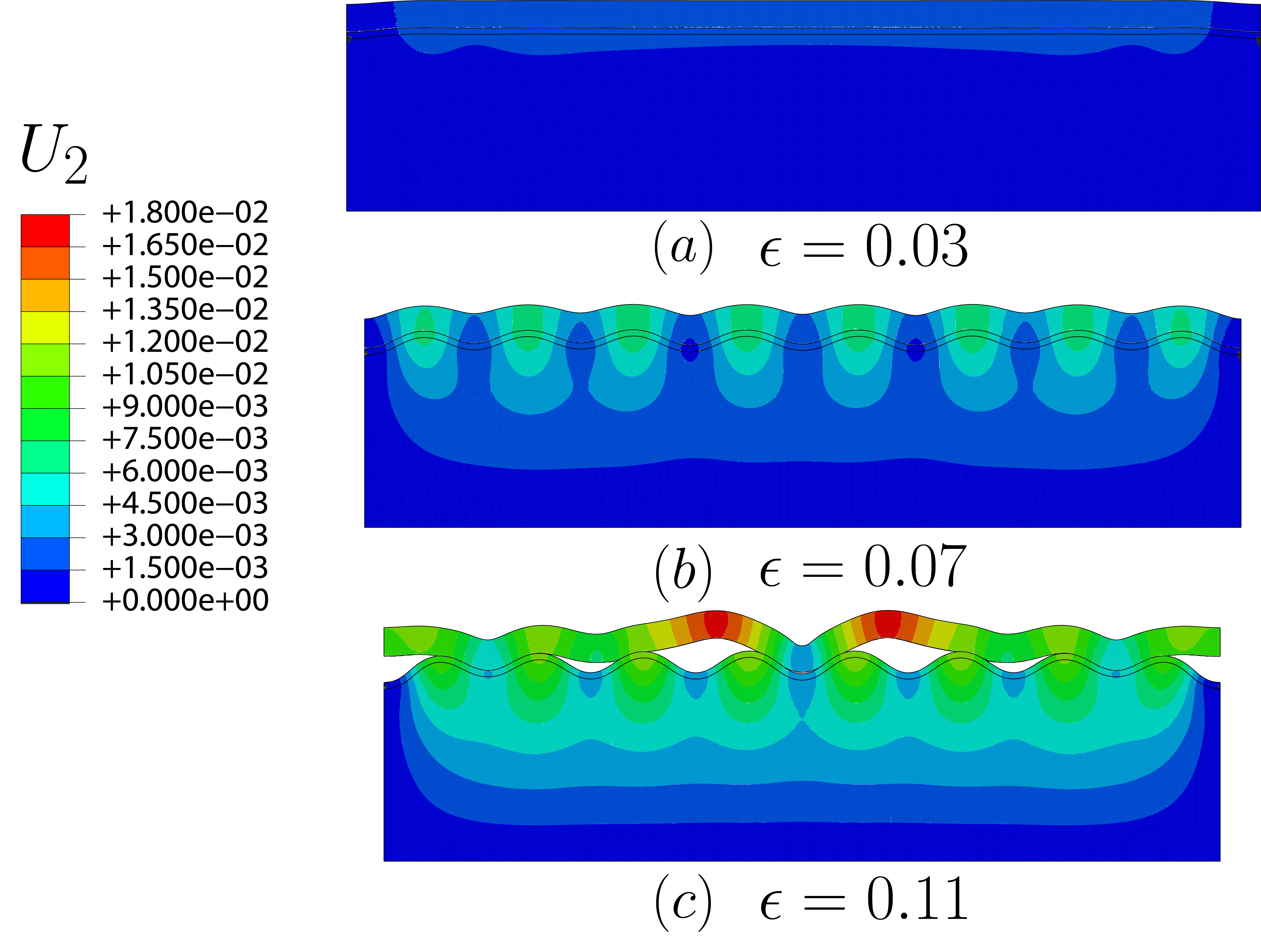}
\caption{Transitions from flat (a) to wrinkled (b) to delamination (c) stages in a typical FE simulation.}
\label{FigS3}
\end{figure*}
\begin{figure*}
\centering
\includegraphics[width=1\textwidth]{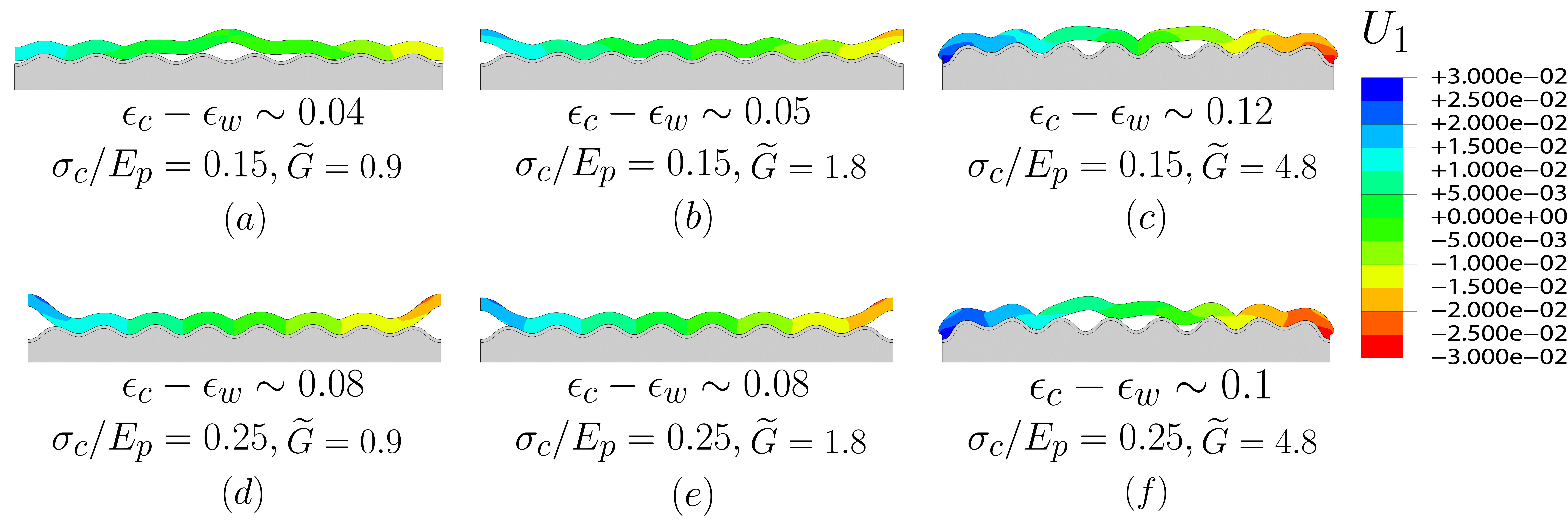}
\caption{FEM simulations showing critical delamination states in the foulant layer for two cases: $\sigma_c/E_p=0.15$ (top) and $\sigma_c/E_p=0.25$ (bottom). Three values of $\widetilde{G}_C$ are used: 0.9 (left), 1.8 (middle), 4.8 (right).}
\label{FigS4}
\end{figure*}
\begin{figure*}
\centering
\includegraphics[width=0.7\textwidth]{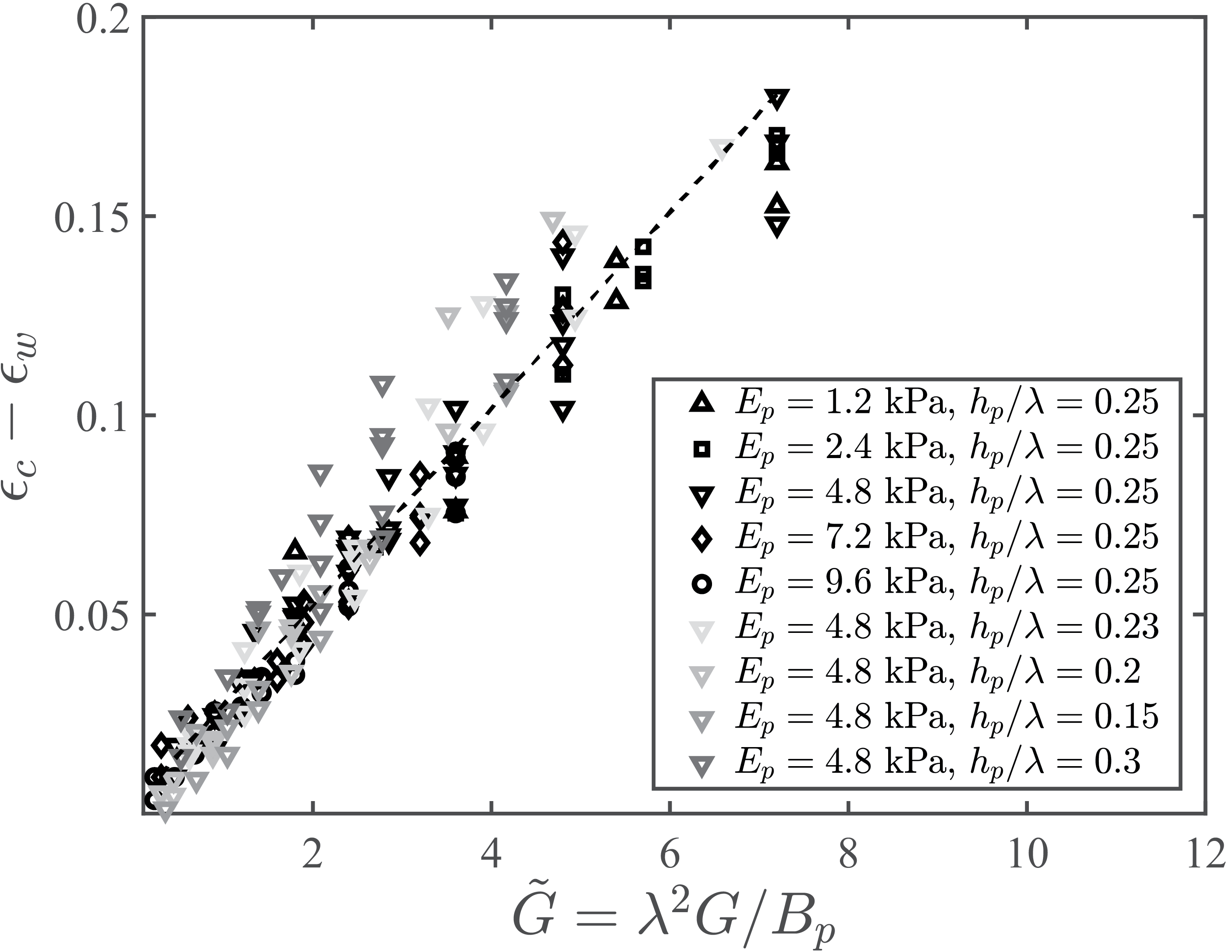}
\caption{$\epsilon_c-\epsilon_w$ with respect to  $\widetilde{G}$ in the energy dominated regime. Various values of $E_p$ and $h_p$ are examined. The black dashed line is a fit showing the linear scaling as predicted from scaling analysis.}
\label{FigS5}
\end{figure*}


%merlin.mbs aipnum4-1.bst 2010-07-25 4.21a (PWD, AO, DPC) hacked
%Control: key (0)
%Control: author (8) initials jnrlst
%Control: editor formatted (1) identically to author
%Control: production of article title (0) allowed
%Control: page (1) range
%Control: year (1) truncated
%Control: production of eprint (0) enabled
\begin{thebibliography}{0}%
\makeatletter
\providecommand \@ifxundefined [1]{%
 \@ifx{#1\undefined}
}%
\providecommand \@ifnum [1]{%
 \ifnum #1\expandafter \@firstoftwo
 \else \expandafter \@secondoftwo
 \fi
}%
\providecommand \@ifx [1]{%
 \ifx #1\expandafter \@firstoftwo
 \else \expandafter \@secondoftwo
 \fi
}%
\providecommand \natexlab [1]{#1}%
\providecommand \enquote  [1]{``#1''}%
\providecommand \bibnamefont  [1]{#1}%
\providecommand \bibfnamefont [1]{#1}%
\providecommand \citenamefont [1]{#1}%
\providecommand \href@noop [0]{\@secondoftwo}%
\providecommand \href [0]{\begingroup \@sanitize@url \@href}%
\providecommand \@href[1]{\@@startlink{#1}\@@href}%
\providecommand \@@href[1]{\endgroup#1\@@endlink}%
\providecommand \@sanitize@url [0]{\catcode `\\12\catcode `\$12\catcode
  `\&12\catcode `\#12\catcode `\^12\catcode `\_12\catcode `\%12\relax}%
\providecommand \@@startlink[1]{}%
\providecommand \@@endlink[0]{}%
\providecommand \url  [0]{\begingroup\@sanitize@url \@url }%
\providecommand \@url [1]{\endgroup\@href {#1}{\urlprefix }}%
\providecommand \urlprefix  [0]{URL }%
\providecommand \Eprint [0]{\href }%
\providecommand \doibase [0]{http://dx.doi.org/}%
\providecommand \selectlanguage [0]{\@gobble}%
\providecommand \bibinfo  [0]{\@secondoftwo}%
\providecommand \bibfield  [0]{\@secondoftwo}%
\providecommand \translation [1]{[#1]}%
\providecommand \BibitemOpen [0]{}%
\providecommand \bibitemStop [0]{}%
\providecommand \bibitemNoStop [0]{.\EOS\space}%
\providecommand \EOS [0]{\spacefactor3000\relax}%
\providecommand \BibitemShut  [1]{\csname bibitem#1\endcsname}%
\let\auto@bib@innerbib\@empty
%</preamble>
\end{thebibliography}%


\begin{thebibliography}{9}

\bibitem{Li2014} Li D, Zheng Q, Wang Y, Chen H. 2014 Combining surface topography with polymer chemistry: exploring new interfacial biological phenomena.
\textit{Polym. Chem.} \textbf{5}, 14--24.

\bibitem{Bixler2012} Bixler GD, Bhushan B. 2012 Biofouling: lessons from nature.
\textit{Phil. Trans. R. Soc. A} \textbf{370}, 2381--2417.

\bibitem{Genzer2006} Genzer J, Groenewold J. 2006 Soft matter with hard skin: from skin wrinkles to templating and material characterization. 
\textit{Soft Matter} \textbf{2}, 310--323.

\bibitem{Pocivavsek2009} Pocivavsek L, Leahy B, Holten-Andersen N, Lin B, Lee KYC, Cerda E. 2009 Geometric tools for complex interfaces: from lung surfactant to the mussel byssus.
\textit{Soft Matter} \textbf{5}, 1963--1968.

\bibitem{Russel2002} Russell T P. 2002 Surface responsive materials.
\textit{Science} \textbf{297}, 964--967.

\bibitem{Shivapooja2013} Shivapooja P, Wang Q, Orihuela B, Rittschof D, L\'{o}pez GP, Zhao X. 2013 Bioinspired surfaces with dynamic topography for active control of biofouling.
\textit{Adv. Mater.} \textbf{25}, 1430--1434.

\bibitem{Levering2014} Levering V, Wang Q, Shivapooja P, Zhao X., L\'{o}pez GP. 2014 Soft robotic concepts in catheter design: an on-demand fouling-release urinary catheter.
\textit{Adv. Healthcare Mater.} \textbf{3}, 1588--1596.

\bibitem{Luka2018} Pocivavsek L, Pugar J, O’Dea R, Ye S, Wagner W, Tzeng E, Velankar S, Cerda E. 2018 Topography-driven surface renewal.
\textit{Nature Physics} \textbf{3}, 948--953.

\bibitem{Luka2019} Pocivavsek  L, Ye S, Pugar J, Tzeng E, Cerda  E, Velankar S, Wagner WR. 2019 Active wrinkles to drive self-cleaning: a strategy for anti-thrombotic surfaces for vascular grafts.
\textit{Biomaterials} \textbf{192}, 226--234.

\bibitem{Nandan2020} Nath NN, Pocivavsek L, Pugar JA, Gao Y, Salem K, Pitre N, McEnaney R, Velankar S, Tzeng E. 2020 Dynamic luminal topography: a potential strategy to prevent vascular graft thrombosis.
\textit{Front. Bioeng. Biotechnol.} \textbf{8}, 573400.

\bibitem{Tindall1988} Svendsen E, Tindall AR.1988 The internal elastic membrane and intimal folds in arteries: important but neglected structures?
\textit{Acta. Physiol. Scand. Suppl.} \textbf{572}, 1--71.

\bibitem{Nguyen2020} Nguyen N, Nath N, Deseri L, Tzeng E, Velankar SS, Pocivavsek L. 2020 Wrinkling instabilities for biologically relevant fiber-reinforced composite materials with a case study of neo-Hookean/Ogden–Gasser–Holzapfel bilayer.
\textit{Biomech. Model. Mechanobiol.} \textbf{19}, 2375--2395.

\bibitem{Hasan2015} Hasan J, Chatterjee K. 2015 Recent advances in engineering topography mediated antibacterial surfaces.
\textit{Nanoscale} \textbf{7}, 15568--15575.

\bibitem{Chen2011} Chen L, Han D, Jiang L. 2011 On improving blood compatibility: from bio-inspired to synthetic design and fabrication of biointerfacial topography at micro/nano scales.
\textit{Colloids Surf. B} \textbf{85}, 2--7.

\bibitem{Mao2009} Mao C, Liang C, Luo W, Bao J, Shen J, Hou  X, Zhao W. 2009 Preparation of lotus-leaf-like polystyrene micro- and nanostructure films and its blood compatibility.
\textit{J. Mater. Chem.} \textbf{19}, 9025--9029.

\bibitem{Koh2010} Koh LB, Rodriguez I, Venkatraman SS. 2010 The effect of topography of polymer surfaces on platelet adhesion.
\textit{Biomaterials} \textbf{31}, 1533--1545.

\bibitem{Shaw2004} Shaw T, Winston M, Rupp CJ, Klapper I, Stoodley P. 2004 Commonality of elastic relaxation times in biofilms.
\textit{Phys. Rev. Lett.} \textbf{93}, 098102.

\bibitem{Christensen2003} Christensen RM. 2003 \textit{Theory of Viscoelasticity}.
New York, NY: Dover Publications.

\bibitem{Wineman2000} Wineman AS, Rajagopal KR. 2000 \textit{Mechanical Response of Polymers}.
Cambridge, UK: Cambridge Univ. Press.

\bibitem{Williams1994} Kinloch AJ, Lau CC, Williams JG. 1994 The peeling of flexible laminates.
\textit{Int. J. Fract.} \textbf{66}, 45--70.

\bibitem{Hutchinson1992} Hutchinson JW, Suo Z. 1992 Mixed mode cracking in layered materials.
\textit{Adv. Appl. Mech.} \textbf{29}, 63--191.

\bibitem{Hutchinson2017} Begley MR, Hutchinson JW. 2017 \textit{The Mechanics and Reliability of Films, Multilayers, and Coatings (Chs 4 and 9)}.
Cambridge, UK: Cambridge Univ. Press.

\bibitem{Vella2009} Vella D, Bico J, Boudaoud A, Roman B, Reis PM. 2009 The macroscopic delamination of thin films from elastic substrates.
\textit{Proc. Natl. Acad. Sci. USA} \textbf{106}, 10901--10906.

\bibitem{Oz2018} Oshri O, Liu. Y, Aizenberg J, Balazs AC. 2018 Delamination of a thin sheet from a soft adhesive Winkler substrate.
\textit{Phys. Rev. E} \textbf{97}, 062803.

\bibitem{Paul2012} Davidson P, Waas AM. 2012 Non-smooth mode I fracture of fibre-reinforced composites: an experimental, numerical and analytical study.
\textit{Phil. Trans. R. Soc. A} \textbf{370}, 1942--1965.

\bibitem{Chai1981} Chai H, Babcock CD, Knauss WG. 1981 One dimensional modelling of failure in laminated plates by delamination buckling.
\textit{Int. J. Solids Struct.} \textbf{17}, 1069--1083.

\bibitem{Roger2013} Chen H, Feng X, Huang Y, Huang Y, Rogers JA. 2013 Experiments and viscoelastic analysis of peel test with patterned strips for application to transfer printing.
\textit{J. Mech. Phys. Solids} \textbf{61}, 1737--1752.

\bibitem{Ravichandran1994} Srinivas MV, Ravichandran G. 1994 Interfacial crack propagation in a thin viscoelastic film bonded to an elastic substrate.
\textit{Int. J. Fract.} \textbf{65}, 31--47.

\bibitem{Thouless2019} Golovin K, Dhyani A, Thouless MD, Tuteja A. 2019 Low–interfacial toughness materials for effective large-scale deicing.
\textit{Science} \textbf{364}, 371--375.

\bibitem{Mei2011} Mei H, Landis CM, Huang R. 2011 Concomitant wrinkling and buckle-delamination of elastic thin films on compliant substrates.
\textit{Mech. Mater.} \textbf{43}, 627--642.

\bibitem{Turon2007} Turon A, Davila CG, Camanho PP, Costa J. 2007 An engineering solution for mesh size effects in the simulation of delamination using cohesive zone models.
\textit{Eng. Fract. Mech.} \textbf{74}, 1665--1682.

\bibitem{Thouless2007} Parmigiani JP, Thouless MD. 2007 The effects of cohesive strength and toughness on mixed-mode delamination of beam-like geometries.
\textit{Eng. Fract. Mech.} \textbf{74}, 2675--2699.

\bibitem{Heinrich2012} Heinrich C, Waas AM. 2012 Investigation of progressive damage and fracture in laminated composites using the smeared crack approach. 
In \textit{53rd AIAA/ASME/ASCE/AHS/ASC Structures, Structural Dynamics and Materials Conference, Honolulu, Hawaii}.

\bibitem{Xie2006} Xie D, Waas AM. 2006 Discrete cohesive zone model for mixed-mode fracture using finite element analysis.
\textit{Eng. Fract. Mech.} \textbf{73}, 1783--1796.

\bibitem{NN2016} Nguyen N, Waas AM. 2016 A novel mixed-mode cohesive formulation for crack growth analysis.
\textit{Compos. Struct.} \textbf{156}, 253--262.

\bibitem{Lin2019} Lin S, Nguyen N, Waas AM. 2019 Application of continuum decohesive finite element to progressive failure analysis of composite materials.
\textit{Compos. Struct.} \textbf{212}, 365--380.

\bibitem{NN2017} Nguyen N, Waas AM. 2017 Continuum decohesive finite element modeling of fiber-reinforced polymer composites: mesh-objectivity and sensitivity studies.
In \textit{58th AIAA/ASCE/AHS/ASC Structures, Structural Dynamics and Materials Conference, Grapevine, Texas}.

\bibitem{Allen1969} Allen HG. 1969 Analysis and Design of Structural Sandwich Panels.
Oxford, UK: Pergamon Press.

\bibitem{Bowden1998} Bowden N, Brittain S, Evans AG, Hutchinson JW, Whitesides GM. 1998 Spontaneous formation of ordered structures in thin ﬁlms of metals supported on an elastomeric polymer.
\textit{Nature} \textbf{393}, 146--149.

\bibitem{Pocivavsek2008} Pocivavsek L, Dellsy R, Kern A, Johnson S, Lin B, Lee KYC, Cerda E. 2008 Stress and fold localization in thin elastic membranes.
\textit{Science} \textbf{320}, 912--916.

\bibitem{Sun2012} Sun J, Xia S, Moon M, Oh KH, Kim K. 2012 Folding wrinkles of a thin stiﬀ layer on a soft substrate.
\textit{Proc. R. Soc. A} \textbf{468}, 932--953.

\bibitem{Cao2012} Cao Y, Hutchinson JW. 2012 Wrinkling phenomena in neo-Hookean ﬁlm/substrate bilayer.
\textit{J. Appl. Mech.} \textbf{79}, 031019.

\bibitem{Cerda2003} Cerda E, Mahadevan L. 2003 Geometry and physics of wrinkling.
\textit{Phys. Rev. Lett.} \textbf{90}, 074302.

\bibitem{Bazant} Bazant ZP, Cedolin L.2010 Stability of Structures: Elastic, Inelastic, Fracture and Damage Theories.
Singapore, SG: World Scientific Publishing,.

\bibitem{Poole2012} Poole RJ. 2012 The Deborah and Weissenberg numbers.
\textit{Rheol. Bull.} \textbf{53}, 32--39.

\bibitem{Abaqus18} Dassault Syst\`emes. 2018 \textit{Abaqus User's Manual, ver. 6.18 [online]}.
MA, USA
\end{thebibliography}
\end{document}